%% file: sim.tex
\documentclass[usenatbib]{mnras}
\include{epsf}
\usepackage{graphicx}
\usepackage{epstopdf}
\input{journals}

\usepackage{color}

\def \farcs{\hbox{$.\!\!^{\prime\prime}$}}

\def \Euclid{\hbox{\it Euclid}}

\title[Sensitivity of shape measurements]{
A study of the sensitivity of shape measurements to the input
parameters of weak lensing image simulations}
\author[Hoekstra et al.]{Henk Hoekstra\thanks{E-mail: hoekstra@strw.leidenuniv.nl}, Massimo Viola, Ricardo Herbonnet\\
  \vspace*{3mm}\\
  Leiden Observatory, Leiden University, PO Box 9513, 2300 RA,
  Leiden, the  Netherlands\\}

\begin{document}

\date{Accepted. Received; in original form}

\maketitle
\begin{abstract}
  Improvements in the accuracy of shape measurements are essential to
  exploit the statistical power of planned imaging surveys that aim to
  constrain cosmological parameters using weak lensing by large-scale
  structure. Although a range of tests can be performed using the
  measurements, the performance of the algorithm can only be
  quantified using simulated images. This yields, however, only
  meaningful results if the simulated images resemble the real
  observations sufficiently well. In this paper we explore the
  sensitivity of the multiplicative bias to the input parameters of
  \Euclid-like image simulations. We find that algorithms will need to
  account for the local density of sources. In particular the impact
  of galaxies below the detection limit warrants further study,
  because magnification changes their number density, resulting in
  correlations between the lensing signal and multiplicative
  bias. Although achieving sub-percent accuracy will require further
  study, we estimate that sufficient archival {\it Hubble} Space
    Telescope data are available to create realistic populations of
  galaxies.
\end{abstract}


\section{Introduction}

In the past decades the theoretical framework that describes the
formation of cosmic structure has been tested by ever more precise
observations \citep[see e.g.][for a comprehensive comparison of
results]{Planck15}, which are in general agreement. However, the main
ingredients of this ``concordance model'' are not understood at all:
dark matter and dark energy make up the bulk, with a mere frosting of
baryonic matter. Although a cosmological constant is an excellent fit
to the current data, its unnaturally small value is by no means
satisfactory. Consequently, many alternative explanations have been
suggested, including modifications of the theory of General Relativity
\citep[see e.g.][for an overview]{Amendola13}. To distinguish between
such a multitude of ideas, dramatically better observational
constraints are needed.

Of particular interest is the study of the distribution of matter as a
function of redshift, because it is sensitive to modified gravity and
the expansion history. The practical complication that most of the
matter is made up of dark matter can be overcome by measuring the
correlations in the ellipticities of distant galaxies that are the
result of the differential deflection of light rays by intervening
structures, a phenomenon called gravitational lensing. The amplitude
of the distortion provides us with a direct measurement of the
gravitational tidal field, which in turn can be used to ``map'' the
distribution of dark matter directly. This makes weak lensing by
large-scale structure, or cosmic shear, one of the most powerful
probes to study dark energy and the growth of structure: we can
determine the statistical properties of the matter distribution as a
function of cosmic time, which depend on the cosmological parameters
\citep[see e.g.][for some recent reviews]{HJ08,Kilbinger15}.

The typical change in ellipticity caused by gravitational lensing is
about a per cent, much smaller than the intrinsic ellipticities of
galaxies. This source of statistical uncertainty can be overcome by
averaging over large numbers of galaxies, although intrinsic
alignments complicate this simple picture \citep[see e.g.][for
reviews]{Joachimi15,Troxel15}. The cosmological lensing signal has now
been measured using ground-based observations of relatively small
areas of sky \citep[e.g.][]{Heymans13, Jarvis16, Jee16, Hildebrandt16b}
but future surveys will cover large fractions of the extragalactic
sky, increasing the source samples accordingly.

The change in ellipticity is also smaller than the typical biases
introduced by instrumental effects. Consequently, averaging the shape
measurements of large ensembles of galaxies is only meaningful if
these sources of bias can be corrected for to a level that renders
them sub-dominant to the statistical uncertainties afforded by the
survey. This will be challenging for the next generation of surveys,
such as \Euclid\footnote{\tt http://euclid-ec.org}~\citep{Laureijs11},
the Wide-Field InfraRed Space Telescope\footnote{\tt
  http://wfirst.gsfc.nasa.gov}~\citep[WFIRST;][]{Spergel15} and the
Large Synoptic Survey Telescope \citep[LSST;][]{LSST09}, which aim to
measure the dark energy parameters with a precision much better than a
percent.

A detailed study of how systematic biases affect the measurements of
galaxy shapes is presented in \cite{Massey13}. This work showed, not
surprisingly, that the PSF is the dominant source, driving the desire
for space-based observations \citep[also see][]{Paulin-Henriksson08}.  Another complication is the fact the
shapes are measured from noisy images. Recent studies have shown that
this leads to biases in the ellipticity
\citep[e.g.][]{Melchior12,Refregier12,Miller13}.  Given a survey
design, our current understanding of these biases, and our ability to
correct for them, requirements can be placed on the instrument
performance, but also on the accuracy of the shape measurement
algorithm. \cite{Cropper13} present a detailed breakdown for \Euclid,
which forms the basis for some of the numbers used in this paper.

Fortunately the impact of the various sources of bias can be studied
by applying the shape measurement algorithm to simulated data, where
the galaxy images are sheared by a known amount. Comparison with the
recovered values then immediately provides a estimate of the bias. For
instance, \cite{Erben01} and \cite{Hoekstra02} used simulated images
to examine the performance of the KSB algorithm developed by
\cite{KSB95}. To benchmark the performance of a wider range of
algorithms, the Shear TEsting Programme \citep[STEP;][]{STEP1,STEP2}
created a blind challenge: the input shear was unknown to the
participants. The results showed a range in performance, even for
algorithms that were in principle rather similar (such as various
implementations of the KSB algorithm). This demonstrated the importance
of how a method is actually implemented. To examine the origin of the
variation in performance further, a series of challenges were carried
out using highly idealized simulations. These Gravitational LEnsing
Accuracy Testing (GREAT) challenges \citep{GREAT08,GREAT10,GREAT3}
have resulted in a steady improvement in the accuracy of the
algorithms given the metric used to compare them, while also
demonstrating the importance of noise on the performance.

However, as recently shown by \citet[][H15 hereafter]{Hoekstra15} the
actual performance of the algorithms depends crucially on the input of
the simulations, such as the distribution of galaxy ellipticities and
the inclusion of faint galaxies. The fidelity of the image simulations
is therefore crucial, not only to quantify biases in the shape
measurements, but also to correctly capture the selection of galaxies
\citep[e.g.][]{FenechConti16}. The aim of this paper is a first
exploration of the sensitivity of shape measurement algorithms to some
of the most basic input parameters in preparation for the next
generation of cosmic shear surveys, and \Euclid\ in particular. This
will help define the range of parameters to consider and to measure
from actual data or simulations.

In \S\ref{sec:shapes} we describe the basic principles of calibrating
a shape measurement pipeline and introduce the algorithm we use. The
image simulations are described in \S\ref{sec:description}. We explore
the sensitivity to the noise level in \S\ref{sec:sim_noise}. The
dependence on the properties of bright galaxies is quantified in
\S\ref{sec:bright}. The impact of faint galaxies is explored in
\S\ref{sec:faint}.  The response to the input ellipticity distribution
is studied in \S\ref{sec:edist} and different implementations of the
algorithm are examined in \S\ref{sec:implementation}. The effect of
stars is evaluated in \S\ref{sec:stardens}.

\section{The need for a calibrated algorithm}\label{sec:shapes}

The measurement of accurate shapes of small, faint galaxies from noisy
data is a critical step in any weak lensing analysis. For this reason
much effort has been focused on reducing the biases in the
measurements of the ellipticity, in particular the correction for the
smearing by the PSF, which leads to rounder images (due to the size of
the PSF) and preferred orientations (if the PSF is
anisotropic). Moreover, all algorithms that measure shapes for
individual galaxies are sensitive to the noise in the images
\citep{Viola14}. Consequently, an ideal algorithm is able to account
for both the biases introduced by the PSF and the noise, because both
tend to vary between exposures. We note, however, that the situation
is complicated further because the object selection itself may lead to
bias: the significance with which galaxies are detected typically
depends on their orientation with respect to the shear or the PSF
\citep{Kaiser00, Bernstein02, Hirata03}. For instance,
\cite{FenechConti16} find that the selection bias can be as important
as the shape measurement bias. Although we do not study selection bias
in this paper, it is clearly another important topic to study in
future work.

The performance of shape measurement algorithms can be estimated using
image simulations. Early comparisons \citep[e.g.,][]{STEP1,STEP2}
included some of the complexity of real data, such as blending of
galaxies. To examine differences between algorithms better, most
recent studies considered idealised circumstances; for instance the
GREAT challenges \citep{GREAT08,GREAT10,GREAT3} focused on isolated
galaxies. However, at the accuracy required for the next generation of
cosmic shear surveys, it is not sufficient to consider such idealised
scenarios as a multitude of subtle effects prevent a straightforward
interpretation of the inferred ellipticity in actual data.

Recently \cite{Huff17} and \cite{Sheldon17} explored an alternative
approach, where the actual survey data are modified, thus avoiding the
use of synthetic data. In this method, which is similar to the one
proposed by \cite{Kaiser00}, the data are sheared by a known amount,
and convolved with additional PSFs to mimic the variation in observing
conditions. The response to these changes provides an estimate of the
multiplicative bias for a particular (biased) shape measurement
method. \cite{Huff17} show how this can reduce the multiplicative bias
for a range of methods. Although these initial results are
encouraging, it is not yet clear whether it is possible to achieve the
stringent requirements for the next generation of surveys, such as
\Euclid.

Sources of bias affect the lensing results in two ways. First of all,
systematics may lead to spurious correlations in the shapes of
galaxies, resulting in an additional signal, i.e. it causes an
additive bias $c$. Although correcting for the various sources of
additive bias may not be easy, residual systematics can typically be
identified by considering cross-correlations between the galaxy shapes
and the source of bias. A well-known example is the star-galaxy
correlation, which is sensitive to residuals in the correction for PSF
anisotropy \citep[see][for a detailed example]{Heymans12}. 

Secondly, the amplitude may be biased by a factor $(1+\mu)$, i.e. 
  the systematics cause a multiplicative bias. The potential of
cosmic shear to constrain dark energy models relies on an accurate
determination of the amplitude of the lensing signal as a function of
source redshift. The amplitude of the lensing signal around
  galaxies as a function of (photometric) redshift can be used to test
  for the presence of multiplicative bias \citep{Velander14}.  Thanks
  to tremendous progress in CMB experiments, comparisons to (future)
  CMB lensing constraints provide an alternative observational test
  \citep[e.g.][]{Liu16,Schaan16}, although the precision may not be
  sufficient. We note, however that these tests are compromised if the
  photometric redshifts themselves are biased. Therefore, unlike
additive biases, the multiplicative bias can only be reliably assessed
through image simulations\footnote{Unless it is a priori known that
  the method is unbiased.}.

The observed shear and true shear are thus related by:

\begin{equation}
\gamma_i^{\rm obs}=(1+\mu)\gamma_i^{\rm true}+c,
\end{equation}

\noindent where we implicitly assumed that the biases are the same for
both shear components. We do so to reduce the number of simulations we
need, but note that in practice the biases need to be determined for
both shear components separately, as we do not expect them to be the
same. Moreover, $\mu$ and $c$ may vary spatially, for instance because
the PSF properties themselves do. In this paper we do not consider
such complications.

Left unaccounted for, multiplicative and additive biases lead to
systematic errors in the inferred cosmological parameters. By
requiring that the systematic shifts in the parameters of interest are
at least smaller than some fraction of the expected statistical
uncertainties from a survey, the maximum allowed range for $\mu$ and
$c$ can be specified. In the case of \Euclid~ the design is based on
the dark energy parameters $w_{\rm p}$ and $w_{\rm a}$, which describe
the constant and dynamic nature of the dark energy, respectively
\citep[see e.g.][]{Laureijs11}. 

Note that the problem is not the amplitude of the bias, but rather how
well the bias can be determined: a known bias can be incorporated as
part of an empirical calibration step, thus reducing the `effective'
residual bias. Hence a robust bias may be preferred over a smaller
bias that is more sensitive to variations in the data or the input
parameters of the simulations: the objective should be to reduce the
sensitivity $|\partial\mu/\partial p|$, where $p$ is a parameter that
may affect the multiplicative bias, for instance the noise level. This
is particularly important if the parameter of interest is correlated
with the lensing signal itself. For example the sensitivity to faint,
undetected galaxies results in correlations with the large-scale
structure that we are trying to measure (see \S\ref{sec:faint} for
more discussion). In some cases the uncertainty in the parameter $p$
may be so large that it leads to an error in the bias that exceeds the
requirement: the method is not calibratable.

Before we continue, it is useful to distinguish between two types of
sensitivity. First of all, methods are sensitive to parameters that
are required to correct for the various sources of bias; incorrect
estimates of these observables lead to biased shape measurements. For
example, the sensitivity to the PSF parameters scales with the ratio
of the square of the PSF size over the galaxy size
\citep{Paulin-Henriksson08,Massey13}.  Similarly, the correction for
noise bias \citep[e.g.][]{FenechConti16} does not reduce the accuracy
with which the noise level needs to be determined.  Hence, the
sensitivity of the algorithm to errors in these parameters needs to be
quantified to establish requirements on how well these parameters need
to be determined from the data.
  
The other kind of sensitivity determines how well the parameters of
interest need to be captured by the image simulations. A shape
measurement algorithm that is truly unbiased\footnote{I.e. it can be
  proven that the algorithm is unbiased in realistic conditions.}
will still be sensitive to errors in the PSF size, noise level,
etc. However, these dependencies can be readily determined by
computing the change in bias when varying the input parameters. Hence
no image simulations are required, because the use of the correct
input parameters is guaranteed to yield an unbiased estimate of the
shear.
  
On the other hand, if a method is biased, image simulations are
required to determine the bias and its sensitivities to the various
input parameters \citep[but see][for an alternative
approach]{Huff17}. The sensitivities can be quantified using
simulations where only the parameters of interest are varied. If we
assume that most effects act independently, as is done in this paper,
this can be done for each parameter separately. Hence the
sensitivities to input parameters can be quantified using a
significantly reduced number of galaxies for which shapes need to be
measured.

Once it has been established that the input parameters are
sufficiently realistic, the actual bias can be determined. The
resulting uncertainty in the bias should be small compared to the
statistical uncertainties from the survey itself, thus defining the
size of the image simulations. In this regard, there is no immediate
advantage to use less biased methods, or even less sensitive methods,
unless the source of bias can be eliminated. However, a lower
sensitivity to a given parameter is clearly preferable because it does
relax the requirements on how well it needs to be captured by the
simulations.

In this paper we examine how the multiplicative bias is affected by
changes in the input parameters of the simulations and by
modifications in the analysis pipeline. We only consider the
multiplicative bias, because it is the most constraining.  It has the
added benefit that we do not have to consider a suite of PSFs with
different ellipticities.

The results presented in \cite{Massey13} suggest a maximum allowed
residual multiplicative bias\footnote{We note that our notation
  corresponds to that of Eqn.\,(11) in \cite{Massey13}, such that
  ${\cal M}\approx 2 \mu$.} of $|\mu_{\rm tot}|<2\times
10^{-3}$. However, as discussed in \cite{Massey13} a number of effects
contribute to this bias, such as errors in the PSF determination and
other corrections for instrumental effects. A detailed discussion of a
possible breakdown is presented in \cite{Cropper13}. We note that
these studies considered requirements under the assumption that
systematic effects do not depend on scale. This can result in
conservative limits, as was discussed in
\cite{Kitching16}. Nonetheless, in order to minimize the
multiplicative bias caused by shortcomings of the image simulations,
we consider an ambitious value of $|\mu_{\rm sim}|=10^{-4}$. We
note that this is not an actual allocation, but rather sets the scope
of the calculations and places requirements on the knowledge of the
input parameters.

To reach a statistical uncertainty of $10^{-4}$ for the
multiplicative bias, a large number galaxies needs to be
analysed. If we consider a shear of $0.01$ and an intrinsic
ellipticity of $0.3$ then a sample of $10^{11}$ galaxies
would be needed. This estimate, however, is too pessimistic, because
the uncertainty is dominated by the intrinsic ellipticity. To reduce
this source of noise, pairs of images, with one rotated by 90 degrees,
can be used \citep[see e.g.][]{STEP2}. The use of more rotations,
e.g., four images rotated by 45 degrees, suppresses shape noise more
efficiently \citep{FenechConti16}, but the performance is ultimately
limited by the pixel noise in the images, such that in practice still
about $10^{10}$ galaxies are required. To ensure that the inferred
biases are robust against the uncertainties in the input parameters,
we wish to explore a range of simulated data. To achieve these
objectives within a reasonable amount of time and limited
computational resources, we thus need to use a sufficiently fast
algorithm.

\subsection{Description of the shape analysis}
\label{sec:analysis}

The impact of (relatively) static sources of bias can be determined
from image simulations, provided they are known and sufficiently well
characterised. However, the instrument configuration varies with time,
as does the atmosphere in the case of ground-based observations; an
ideal shape measurement method should therefore accurately correct for
the resulting temporal variations in the PSF in order to avoid having
to create a large suite of simulations for all possible PSFs. In this
paper we do not consider varying PSFs, but assume that this can
be accurately corrected for. Furthermore, we do not consider the
impact of selection effects, which will be important in realistic
situations, as shown by \cite{FenechConti16}. Instead we focus on the
sensitivity of the shape measurements to the input parameters of the
image simulations.

Given an image and a model for the PSF (we assume that other sources
of bias have been removed to a sufficient level of accuracy) different
approaches can be used to estimate the true galaxy shape. For
instance, one can fit a parametrised model of the surface brightness
distribution to the data by representing galaxies by a decomposition
into shapelets \citep{Refregier03}. The resulting model can then be
deconvolved analytically to yield an estimate of the underlying galaxy
shape. However, the expansion into shapelets needs to be truncated
because of noise.

To avoid the problems with direct deconvolution, forward modeling
techniques have become more popular. In this case a model image is
sheared, convolved with the PSF and pixellated. The model parameters
are varied until a best fit to the data is obtained. This step
requires many calculations, especially if more model parameters are to
be determined. Accurate priors for the parameters are required to
obtain an unbiased estimate for the shear. These priors can be
  derived from high-quality observations, similar to what is needed
  as input for the image simulations. Moreover, the data themselves
  can be used to update the priors. Analogous to the need to
truncate the shapelet expansion, the model should provide a good
description of the galaxies, while having a limited number of
parameters in order to avoid over-fitting. A model that is too rigid
will lead to model bias \citep[e.g.][]{Bernstein10}, whereas a model
that is too flexible tends to fit noise in the images
\citep[e.g.][]{Kacprzak12}. We note that this can mitigated by
a marginalisation of the nuisance parameters in the 
model \citep{Miller13}.

Unfortunately, fitting methods require many evaluations, thus increasing
the computational time per galaxy significantly. We focus instead
on an alternative approach: we measure the moments of
the galaxy images, which are subsequently corrected for the PSF. The
shapes are quantified by the polarization\footnote{We define the
  ellipticity $\epsilon\equiv (a-b)/(a+b)$, with $a$ and $b$ the major
  and minor axes, respectively. The polarization $e$, for such a
  galaxy would be approximately $(a^2-b^2)/(a^2+b^2)$}:

\begin{equation}
e_1=\frac{I_{11}-I_{22}}{I_{11}+I_{22}},~{\rm and~}
e_2=\frac{2I_{12}}{I_{11}+I_{22}},\label{eq:polarisation}
\end{equation}

\noindent where the quadrupole moments $I_{ij}$ are given by

\begin{equation}
I_{ij}=\frac{1}{I_0}\int{\rm d}^2{\bf x}\, x_i\, x_j\, W({\bf x})\, f({\bf x}),
\end{equation}

\noindent where $f({\bf x})$ is the observed galaxy image,
$W({\bf x})$ a suitable weight function to suppress the noise and
$I_0$ the weighted monopole moment. 

Matching the width of the weight function to the object size maximizes
the signal-to-noise of the shape measurement. For the weight function
we adopt a Gaussian with a dispersion $r_{\rm g}$ determined by the
half-light radius $r_{\rm h}$ measured by {\tt SExtractor}
\citep{Bertin96}.  As reference we consider
$r_{\rm g}=r_{\rm h}/\sqrt{2}$, which is slightly smaller than the
optimal value for a Gaussian, which would imply
$r_{\rm g}=r_{\rm h}/\sqrt{2\ln 2}$. This choice does not affect our
our conclusions, and we explore different choices in
\S\ref{sec:implementation}.  A further sophistication would be to try
to match the shape of the weight function
\citep[e.g.][]{Melchior11,Okura11}. This optimization slows down the shape
measurement algorithm and increases the sensitivity to the input
ellipticity distribution. As the gain is expected to be small for the
smallest galaxies, we limit our study to an axisymmetric weight
function.

In practice an estimate for the background needs to be subtracted from
the observed image and contributions from nearby objects
suppressed. In our reference setup we simply mask pixels within a
radius of $4 \, r_{\rm g}$ around neighbouring objects, and the
background is determined locally by considering an annulus with inner
and outer radii of 16 and 32 pixels, respectively, from the centroid
of the object. Objects located in the annulus are masked and a plane
is fit to the counts in the unmasked pixels. We prefer a local
background determination over a global one, because biases due to
artifacts are limited to relatively small scales, and thus do not
introduce coherent biases on scales that are relevant for the
cosmological interpretation. In \S\ref{sec:implementation} we explore
different settings, demonstrating that the background determination is
an essential aspect of the algorithm performance.

The resulting weighted moments are biased because of the weight
function and PSF. To undo these we focus here on the commonly used KSB
method developed by \cite{KSB95} and \cite{LK97} with corrections
provided in \cite{Hoekstra98} and \cite{Hoekstra00}, which was used in
H15. The only difference with H15 is that we use {\tt SExtractor}
for the object detection step to speed up the analysis. We note that
the KSB algorithm makes simplifying assumptions about the PSF which
are not valid for realistic cases, such as the \Euclid~ PSF. However,
this can be accounted for with improved moment-based methods such as
{\tt DEIMOS} \citep{Melchior11}.

We stress, however, that the aim of this paper is not to find a
suitable shape measurement algorithm, nor are we interested in the
value of the bias. Instead we explore the change in bias as a function
of the input parameters of the image simulations. Exactly this crucial
step has been largely overlooked in previous work. However, as we will
see, most of the variations in bias are small and therefore not that
important for current surveys. On the other hand, our results also
clearly demonstrate that the situation is fundamentally different for
\Euclid, because of the much more stringent requirements.

\section{Description of the image simulations}
\label{sec:description}

A flexible framework to create simulated images is provided by {\tt
  GalSim} \citep{Rowe15}, a publicly available code that was developed
for GREAT3 \citep{Mandelbaum14,GREAT3}. Here we use simple parametric
models for the galaxies, and the main input is a list of galaxy
properties with a position, flux, half-light radius, Sercic index and
ellipticity, from which sheared galaxy images are computed. The reference
galaxy parameters are described in more detail in \S\ref{sec:input}.

As input to the simulations we use a sample of galaxies for which
morphological parameters were measured from resolved $F606W$ images
from the GEMS survey \citep{Rix04}. These galaxies were modeled as
single Sersic models with {\tt GALFIT} \citep{Peng02} and for our
study we use the measured half-light radius, apparent magnitude and
Sersic index $n$. For simplicity, we only consider galaxies fainter
than magnitude $m=20$. 

\begin{figure}
\centering
\leavevmode \hbox{%
\includegraphics[width=8.5cm]{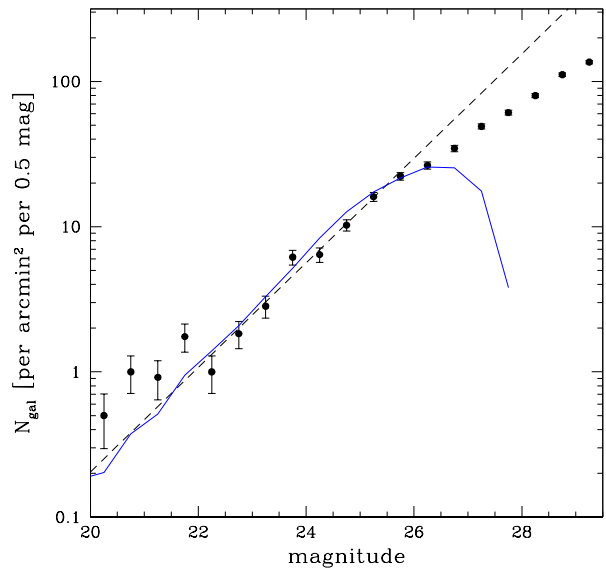}}
\caption{Number density of galaxies as a function of limiting
magnitude. The dashed black line indicates our reference model,
which is a power law with a slope of $0.36$, which is a good
description of the GEMS counts (solid blue line) for $20<m<25$.
The points with errorbars are the $F775W$ counts from \citet{Coe06} based
on the UDF, which suggest a flatter slope at faint magnitudes.
\label{fig:input_counts}}
\end{figure}

The solid blue line in Fig.~\ref{fig:input_counts} shows the number
density of galaxies in the GEMS catalog as a function of apparent
magnitude. For $20<m<25$ the counts are well described by a power law
with a slope of 0.36, indicated by the black dashed line. Comparison
with the observed $F775W$ counts from the Hubble Ultra Deep Field (UDF) by
\cite{Coe06} shows that for $m>26$ the GEMS catalog is incomplete.
Although the UDF counts indicate a flattening of the slope for
$m>25.5$, our reference simulations assume a simple power law with
slope 0.36, but in \S\ref{sec:slope} we explore the sensitivity of the
results to this assumption.

The input catalog is normalised to 36 galaxies arcmin$^{-2}$ with
$20<m<24.5$. We choose a nominal noise level per pixel with a
dispersion of 0.8, which results in a typical number density of 47
galaxies arcmin$^{-2}$ with a signal-to-noise ratio larger than 10, as
measured by {\tt SExtractor}, and a number density of 33 galaxies
arcmin$^{-2}$ if we restrict the magnitude range to $20<m<24.5$.  We
measure shapes for all detected galaxies, but report results for
galaxies with $20<m<24.5$ and $r_{\rm h}>0\farcs11$ unless specified
otherwise. For this magnitude range, the cut in half-light radius
cleanly separates galaxies from stars. We note that these number
densities are somewhat higher than the nominal values for \Euclid\
\citep{Laureijs11}.

We create pairs of images where the galaxies are rotated by 90 degrees
in the second image in order to reduce the noise due to the intrinsic
ellipticity distribution \citep[see e.g.][]{STEP2}: by construction
the mean intrinsic ellipticity when both are combined is zero.  We
analyse the images separately and thus, due to noise in the images,
this is no longer exactly true, especially for faint galaxies. The
input shears range from $-0.06$ to 0.06 in steps of 0.01 (for both
components), yielding 169 image pairs for each ``set''.  We use the
{\tt galsim.applyShear()} function\footnote{In version 1.1 this
  method was deprecated.}, which preserves the area of the object,
i.e. the galaxies are not magnified.  We verified this by measuring
the average sizes of the galaxies as a function of the applied
shear\footnote{We found that the mean observed half-light radius
  increased by a negligible 0.2\% for the largest shear we consider
  here.  However, closer investigation revealed that this change in
  size is solely due to a direction-dependent feature in the way the
  half-light radius is determined.}. This greatly simplifies the
interpretation of our results, as magnification changes the galaxy
selection as a function of shear. Although not the focus of this
paper, we discuss the implications of our results on size
magnification studies in \S\ref{sec:magnification}.

Each image has a size of 10\,000 by 10\,000 pixels, with a pixel scale
of $0\farcs 10$, corresponding to that of the {\it Euclid} VIS camera. Down to a
limit $m=29$ (see \S\ref{sec:faint}) for each pair of images we
include $10^6$ objects, or $1.7\times 10^8$ per set.  To reduce the
statistical uncertainties further we simulate typically tens of sets
created with different random seeds. The analysis of a single set,
using the reference setup of the KSB algorithm, takes approximately 60
core hours on a Dell PowerEdge R820 with Intel Xeon E5-4620 2.20GHz
processors. For example, the results presented in
Fig.~\ref{fig:bias_mlim} took over 34\,000 core hours.

To simulate a diffraction-limited telescope in space, we adopt a
circular Airy PSF for a telescope with a diameter of 1.2m and a PSF
obscuration of 0.3 at a reference wavelength of 800nm. This is a
reasonable approximation to the \Euclid~ PSF in the VIS-band. We include a small
number of bright stars in the simulations, which are used to measure
the PSF properties required to correct the galaxy shapes. In this
paper we do not consider the complications that arise from modeling of
the PSF. In \S\ref{sec:stardens} we do explore the impact of
variations in the star density on the multiplicative bias.

\subsection{Input galaxy properties}
\label{sec:input}

The morphological properties, such as sizes, shapes and surface
brightness profiles are correlated: fainter galaxies are on average
smaller, whereas disk-dominated galaxies show a broader ellipticity
distribution. To capture some of these correlations we use
measurements of morphological parameters (specifically magnitude,
observed half-light radius and Sersic index) from the resolved $F606W$ images
from the GEMS survey \citep{Rix04}. The use of Sersic profiles to
describe the galaxies limits the fidelity of the simulations
\citep[see][for a study of the biases that may arise]{Kacprzak14}, and
future work will need to examine how well morphological parameters
need to be determined, including the spatial variation of galaxy
colors \citep{Semboloni13}.

Although \cite{Rix04} also estimated ellipticities, we ignore any
correlation with ellipticity, but randomly draw ellipticity values
from a Rayleigh distribution given by

\begin{equation}
P(\epsilon;\epsilon_0)=\frac{\epsilon}{\epsilon_0^2}\, {\rm e}^{-\epsilon^2/
2\epsilon_0^2},
\end{equation}

\noindent where the value of $\epsilon_0$ determines the width of the
distribution, as well as the average
$\langle\epsilon\rangle=\epsilon_0 \sqrt{\pi/2}$. We need to truncate
the distribution because the ellipticity cannot exceed unity, but also
because galaxy disks have a finite thickness. We therefore set
$P(\epsilon,\epsilon_0)=0$ if $\epsilon>0.9$. 

As reference value we adopt $\epsilon_0=0.25$ which best described the
data used in H15. This simplyfing assumption, i.e. that $P(\epsilon)$
is independent of other galaxy properties, will need to be studied in
future work, as the ellipticity distributions for early and late type
galaxies differ \citep[e.g.][]{Uitert11}, which in turn results in
dependencies with the environment \citep{Kannawadi15}. In practice the
ellipticity distributions for various subsets will need to be measured
from deep observations \citep{Viola14}.

We consider bins with a width of 0.1 in magnitude and compute the
expected number of galaxies assuming a power law for the galaxy counts
as described above. Down to a limiting magnitude of $m=25.4$ we
randomly draw galaxies from the corresponding magnitude bin from the
GEMS catalog. For fainter galaxies we create duplicates with different
orientations and place those postage stamps in the image. This speeds
up the creation of the simulated images. 

\subsubsection{Sizes of faint galaxies}

H15 showed that a robust estimate of the multiplicative bias requires
the inclusion of sufficiently faint galaxies in the image simulations:
for current ground-based observations the bias converges when galaxies
that are 1.5 magnitude fainter than the faintest galaxy used in the
lensing analysis are present.  As discussed in more detail in
\S\ref{sec:faint} in the case of \Euclid~ we may need to consider
galaxies as faint as magnitude 29.

For galaxies brighter than $m=26.5$ we use the observed
half-light radii from \cite{Rix04} instead of the {\tt GALFIT}
estimate for the effective radius, because it is a more robust
estimate\footnote{These are the values measured by SExtractor,
  and thus not corrected for the HST PSF. This omission, which we
  discovered during the refereeing process, slightly biases the sizes
  used in our analysis: at $m=24.5$ the corrected sizes are 4\%
  smaller and the differences are even smaller for brighter galaxies.
  The change in size is larger for fainter galaxies, but this only
  affects the recovered biases but not the sensitivities themselves
  (see e.g. Fig.~\ref{fig:bias_size}).}. However, for $m>26.5$ the
GEMS catalog becomes progressively incomplete, resulting in a biased
distribution of galaxy sizes.  Given the small space-based PSF and the
stringent requirements of the \Euclid~ mission the adopted
distribution of galaxy sizes may be relevant. This is quantified in
\S\ref{sec:galsize} but here we describe how we parametrized the size
distribution of the faint galaxies.

\begin{figure}
\centering
\leavevmode \hbox{%
\includegraphics[width=8.5cm]{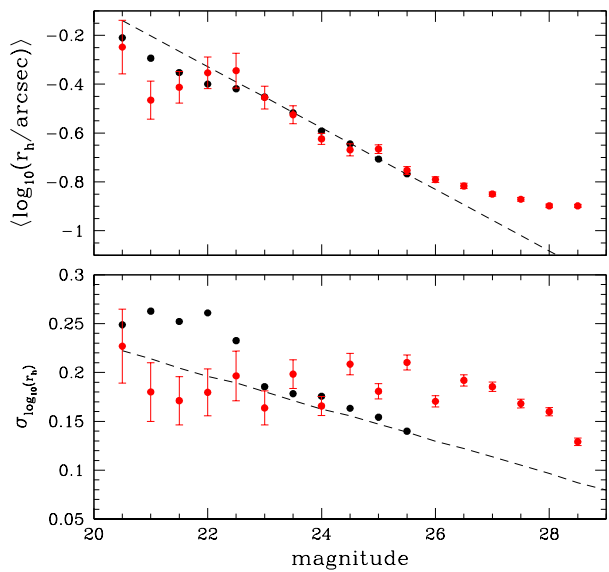}}
\caption{{\it Top panel:} mean logarithm of the observed half-light
  radius as a function of apparent magnitude. The black points
  indicate the measurements from GEMS that are used to derive our
  parametric model (indicated by the dashed line; see text for
  details). The red points indicate the results from the UDF,
  suggesting that the actual sizes of faint galaxies may be larger.
  {\it Bottom panel:} width of the galaxy size distribution as a
  function of limiting magnitude, where the black points correspond to
  GEMS and the red points are from the UDF. The dashed line shows the
  value for the parametric model adopted for the image simulations.
\label{fig:model_size}}
\end{figure}

As shown in Appendix~\ref{app:size} the distribution of observed half-light
radii for bright galaxies can be approximated by a skewed log-normal
distribution. We keep the skewness fixed to a representative value of
$-0.58$ and determine the mean and dispersion of
$\log_{\rm 10} r_{\rm h}$ as a function of apparent magnitude. The
results are presented as the black points in
Fig.~\ref{fig:model_size}.  For galaxies with $23<m<25.5$ both the
mean and the dispersion of $\log_{\rm 10} r_{\rm h}$ are well
described by a simple linear relation. We determine the best fit in
this magnitude range (dashed lines) and use this to describe the size
distribution for galaxies with $m>26.5$.

\cite{Coe06} provide estimates for the effective radii, $r_{\rm eff}$,
from the best fit {\tt GALFIT} model using HST observations
of the HST UDF. To allow for a direct comparison to our input
parameters, we convert the values for $r_{\rm eff}$ to half-light
radii using an empirical relation based on the GEMS catalog, which
provides both size estimates (see Appendix~\ref{app:size} for
details). We only consider galaxies for which the effective radius was
determined with a relative precision $<10\%$ and present the results
in the top panel of Fig.~\ref{fig:model_size} (red points). The
agreement with the GEMS meausurements and our parametric model is good
for $m<26$. Interestingly, the results from \cite{Coe06} suggest that
the sizes of the faint galaxies may be larger than we assume here.

Similarly, we use the GEMS results to relate the dispersion in
$r_{\rm eff}$ to an estimate of the scatter in $r_{\rm h}$. As shown in the
bottom panel of Fig.~\ref{fig:model_size}, the resulting converted
measurements from \cite{Coe06} roughly match our adopted
model. Although this level of agreement is adequate for the purpose of
this paper, it will clearly be worthwhile to revisit the size measurements
presented in \cite{Coe06}.

\section{Sensitivity to noise}
\label{sec:sim_noise}

As shown in \cite{Viola14} any ellipticity measurement is biased in
the presence of noise in the image, because the estimator is
non-linear in terms of the pixel values: Eqn.~(\ref{eq:polarisation})
involves a ratio of moments. As a consequence the observed ellipticity
distribution is skewed and not centered on the true value. This needs
to be accounted for when estimating the shear from an ensemble of
sources, because it otherwise leads to a multiplicative bias.

In the actual observations the background level is expected to vary:
in ground-based data due to the moon or changing atmospheric
conditions and in space-based observations because of the zodiacal
background which varies across the sky. Consequently, the shape
measurement algorithm needs to be able to account for the dependence
on the signal-to-noise ratio
\citep[SNR;][]{Miller13,Hoekstra15,FenechConti16}, which in turn
implies that the statistics of the background need to be determined
sufficiently well. This is typically done by measuring the
distribution of unmasked pixel values where no galaxies were detected,
i.e assuming that the noise is uncorrelated. We turn to the
complications posed by faint undetected galaxies in \S\ref{sec:faint},
and consider a simple background first.

\begin{figure}
\centering
\leavevmode \hbox{%
\includegraphics[width=8.5cm]{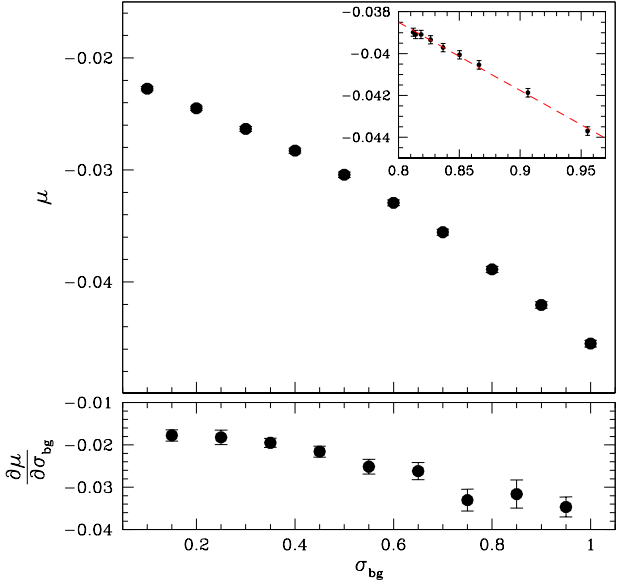}}
\caption{Multiplicative bias for galaxies with $20<m<24.5$ 
as a function of $\sigma_{\rm bg}$, the rms of the background.
The errorbars indicate the dispersion in the results. The bias increases with
increasing noise level and the slope  $d\mu/d\sigma_{\rm bg}$ steepens, as is 
evident from the bottom panel. For our implementation of 
the KSB algorithm we find that the bias around $\sigma_{\rm bg}=0.8$ 
(see inset in the top panel) can be approximated by a linear relation 
with a slope  $\partial\mu/\partial\sigma_{\rm bg}=-0.0326\pm0.0007$.
\label{fig:bias_noise}}
\end{figure}

To do so, we simulate 20 sets of images where we include galaxies down
to a limiting magnitude of $m_{\rm lim}=24.5$. We add Gaussian noise
with a dispersion $\sigma_{\rm bg}$ (the images are created with a
mean background of zero). Note that we are not attempting to simulate
the actual observing process, which involves the combination of
multiple exposures. To reduce the number of simulations the seeds for
the noise realisations and galaxy properties are the same between
sets. The images are analysed as described in \S\ref{sec:analysis}.
The resulting observed multiplicative bias $\mu$ as a function of the
background level is presented in the top panel of
Fig.~\ref{fig:bias_noise}. As expected the bias increases when the
background noise level is higher. Also note that $\mu$ does not vanish in
the absence of noise, a demonstration of the fundamental limitation of
using the KSB algorithm. The inset panel shows the change in bias
close to the nominal background $\sigma_{\rm bg}=0.8$ that we use for
the other results presented in this paper.

The bottom panel of Fig.~\ref{fig:bias_noise} shows the slope
$\partial\mu/\partial \sigma_{\rm bg}$ which steepens as the noise
level increases. When considering a small range, such as the
one shown in the inset of the top panel, a constant slope is a 
good approximation and around $\sigma_{\rm bg}=0.8$ we measure
a value $\partial\mu/\partial \sigma_{\rm bg}=-0.0326\pm0.0007$. 

We can use this result to estimate how well the background rms needs
to be determined, given an allocation for the bias that can be
tolerated: a maximum uncertainty of $\delta\mu=10^{-4}$ implies that
$\sigma_{\rm bg}$ needs to be measured with a relative precision of
approximately $0.3\%$. If the noise is homoscedastic and Gaussian, as we
assumed here, this requires about $2\times 10^5$ blank pixels.  In
practice undetected cosmic rays, galaxies, flat-field errors, etc. may
also contribute to the background statistics. These are naturally
included when blank pixels are used to characterize the background,
provided instrumental effects that modify the statistics of the
observed background do not do so over the area that contains
$2\times 10^5$ blank pixels. If we assume that half the pixels
are blank, this corresponds to a square patch of 650 pixels on a side,
which is much smaller than the section that is controlled by a single
component of read-out electronics.

We stress that we are only concerned with coherent errors in the
determination of background statistics as these lead to bias on
cosmologically interesting scales. Small scale effects that do not
correlate between detectors or exposures merely increase the
measurement noise slightly, which is negligible compared to the
intrinsic shape noise on such small scales \citep[see e.g.][]{Kitching16}.

\section{Properties of bright galaxies}
\label{sec:bright}

The presence of noise in the images prevents us from using unweighted
moments, for which the correction for the smearing by the PSF is
trivial. To relate the observed weighted moments to the true
(unweighted) moments requires estimates of the higher order moments or
equivalently the morphology of the galaxies
\citep[e.g.][]{Semboloni13}.  Only noisy estimates can be obtained
from the data, and thus the bias depends not only on the noise level,
but also on the underlying distribution of galaxy morphologies. This
remains an important open topic of study, albeit not as prominent as
the two properties that we will study in this section: the sizes and
number densities of galaxies.

\begin{figure}
\centering
\includegraphics[width=8.5cm]{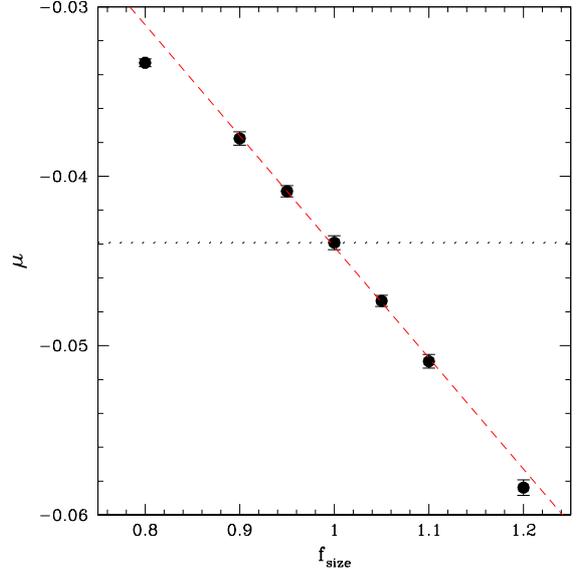}
\caption{Change in multiplicative bias for galaxies with
    $20<m<24.5$ when the sizes of the input galaxies (we only
    include galaxies brighter than $m_{\rm lim}=26$ in the
    simulations) are increased by a factor $f_{\rm size}$. For
  reference, the hatched region indicates a variation of $10^{-4}$ in
  the value of $\mu$. The red dashed line is the best-fit linear
relation between $\mu$ and $f_{\rm size}$.
\label{fig:bright_size}}
\end{figure}

\subsection{Sensitivity to galaxy size}

The larger a galaxy, relative to the PSF, the easier it should be to
measure its shape. Consequently, the multiplicative bias is typically
a (strong) function of the galaxy size.  To quantify the sensitivity
to input galaxy size, we make simulated images where we change the
sizes of the input galaxies by a factor $f_{\rm size}$. We created 10
sets for each value of $f_{\rm size}$, where we include galaxies down
to a limiting magnitude $m_{\rm lim}=26$. We measure $\mu$ for
galaxies with $20<m<24.5$ and show the results in
Fig.~\ref{fig:bright_size}. The sensitivity is substantial, with the
dashed line indicating the best fit to the measurements for
$0.9\leq f_{\rm size}\leq 1.1$. Note that the slope is so steep
because for our implementation $\mu$ is a strong function of galaxy
size.  We find a slope
$\partial\mu/\partial f_{\rm size}=-0.0656\pm 0.0010$, which would
imply that $f_{\rm size}$ should be determined with a precision of
0.15\% if we consider a maximum tolerance of $\delta\mu=10^{-4}$.

This is unnecessarily conservative, because it would be apparent from
the data themselves that the mean sizes differ from the
simulations. This could then be corrected by improving the realism of
the image simulations. For instance, \cite{Bruderer16} propose the use
of Monte Carlo Control Loops to adjust the simulated data such that
they are statistically consistent with the observations. An
alternative approach was explored in \cite{FenechConti16} who
resampled the output from the simulations to match the size
distribution in the data. Although this provides an improved estimate
for the bias for the observed sample of galaxies, local variations in
galaxy sizes will not be captured. 

It is therefore preferable to account for any size dependence in the
algorithm through an empirical calibration. Note that such a
calibration, or `training', is not restricted to size, but can include
any parameter of interest, such as brightness, surface brightness
profile, local environment, etc. In this case the image simulations
are used to identify correlations between multiplicative bias and
observables. For instance \cite{Tewes12} explored the use of
supervised machine learning.  Similarly \cite{Gruen10} used a neural
network to remove residual biases. Interestingly, the run time is not
determined by the application of the trained algorithm to real data,
but rather by the time it takes to analyse the simulated
data. Compared to typical machine learning applications, the training
sample is much larger than the actual data sample, because we wish to
reduce the uncertainty in multiplicative biases by considering a very
large volume of image simulations. Importantly, the fidelity of the
machine learning step depends critically on using appropriate
inputs. As we discuss in \S\ref{sec:faint}, this includes capturing
the impact of galaxies below the detection limit.

Inevitably, the empirical corrections are based on observed parameters
that are noisy. As a consequence the calibration may be biased if the
input properties are incorrect. Moreover, the choice of size
definition matters. This was highlighted in \cite{FenechConti16} who
showed how selection biases in both the detection and analysis steps
may result in implicit selections in ellipticity, and consequently
lead to biases in the recovered shear. Hence, particular care should
be taken in ensuring that consistent size definitions are used. We do
not study the impact of these selection effects here, but stress that
these represent an important source of bias, especially when
selections are made using parameters that correlate strongly with
ellipticity. Moreover, as such selection biases appear inevitable in
the presence of PSF anisotropy and shear, it remains unclear whether
image simulations covering a wide range of instrument states can be
avoided.

\subsection{Sensitivity to galaxy density}
\label{sec:galdens}

Close pairs of galaxies are another complication in real data. Most
recent work on the performance of shape measurement algorithms focused
on isolated galaxies: in \cite{GREAT08}, \cite{GREAT10}, \cite{GREAT3}
and \cite{Jarvis16} only postage stamps of isolated galaxies were
analyzed, and thus the effects of blending were not included. The
image simulations we study here do include close pairs, but do not
capture the full complexity expected in real data as galaxies are not
positioned randomly on the sky, but are instead clustered. Hence the
local density of galaxies varies significantly, with clusters of
galaxies representing the most extreme cases. For instance, for a
sample of massive clusters at $z\sim0.2$ H15 find that the number
density of the brightest galaxies ($20<m<21$) is on average increased
by a factor two at a radius of 1~Mpc, whereas the number density of
fainter galaxies ($24<m<25$) is increased by approximately $20\%$.

One of the objectives of the \Euclid\ mission is to use the number
density of galaxy clusters as a function of mass and redshift to
constrain cosmological parameters \citep{Sartoris15}, which relies on
accurate mass estimates. As shown by \cite{Koehlinger15} this should
be possible, provided that the multiplicative bias for this application is
comparable to the required accuracy for cosmic shear studies. 

\begin{figure}
\centering
\includegraphics[width=8.5cm]{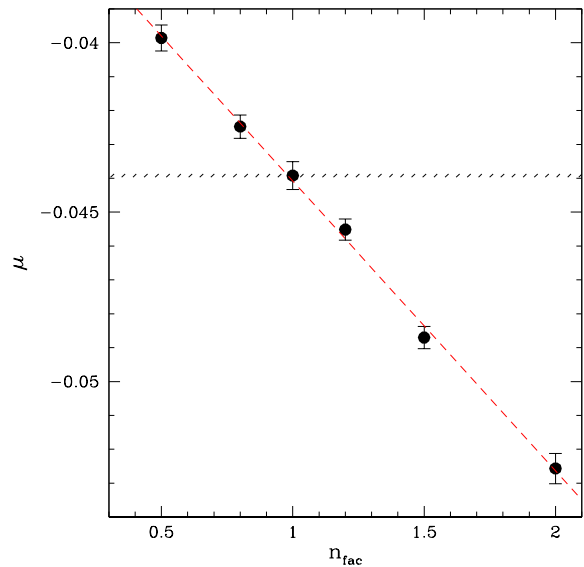}
\caption{Multiplicative bias for galaxies with magnitude
$20<m<24.5$ as function of $n_{\rm fac}$, the relative increase in the number
density of galaxies brigher than magnitude 26. For reference, the
hatched region indicates a variation in $\mu$ of size $10^{-4}$.
\label{fig:bias_nfac}}
\end{figure}

Given the impact of blending on shape measurements, it is useful to
examine how $\mu$ depends on the number density of bright galaxies.
To do so we simulated observations with $m_{\rm lim}=26$, while
increasing the number density by a factor $n_{\rm fac}$ compared to
the reference case. We measure $\mu$ for galaxies with $20<m<24.5$ and
show the results in Fig.~\ref{fig:bias_nfac}. The red dashed line
indicates the best fit linear relation with a slope
$\partial\mu/\partial n_{\rm fac}=-0.00856\pm0.00017$. These results
show a strong (linear) dependence on the local number density of
bright galaxies. If unaccounted for, this will lead to significant
biases in cluster mass estimates from \Euclid.

The dependence on the number density of bright galaxies does not only
affect cluster studies, but also complicates the cosmic shear
analysis. The large-scale structure that gives rise to the signal
(with most of the contribution coming from halos that correspond to
galaxy groups) is correlated with the density of galaxies. Hence the
multiplicative bias is coupled to the cosmic shear signal and the
image simulations should thus capture the clustering of galaxies well.
This is a concern, because the predicted clustering signal itself
depends on cosmology. 

We should therefore strive for algorithms that are minimally sensitive
to neighbouring objects. Our deblending implementation is simple and
more sophisticated approaches will be studied in future work to
quantify the impact of clustering.  For instance, we note that
\cite{FenechConti16} measured a change of $2\times 10^{-3}$ in
multiplicative bias for their simulations of ground based data when
the number density of galaxies was reduced by a factor 2.  We do
expect fitting methods to perform better than our KSB implementation,
because the estimates of the moments are biased by simply masking
blended objects. In contrast, fitting methods are naturally less
sensitive to masked areas, but can also be adapted to fit multiple
galaxies simultaneously. Whether this can fully eliminate the effect
of blending should be studied further.

As a first exploration for moment-based methods we analysed a smaller
set of image simulations using two alternative implementations (see
\S\ref{sec:implementation} for more variations). First, we switched
off the masking of neighbouring detected galaxies. In this case the
bias increases, and the slope $\partial\mu/\partial n_{\rm fac}$ is
30\% steeper. As blending may affect the local background
determination, we also considered the case where the background is
fixed to 0, the correct level in the simulated data. In this case the
bias is indeed reduced, but the slope is unchanged. These findings
demonstrate that differences in implementation play a role, but appear
unable to significantly reduce the sensitivity to blending. Something
to investigate in future work is whether the impact of blending can be
alleviated by interpolating over the masked regions.

Alternatively, the bias can be determined as a function of both
distance to the nearest neighbour and its flux difference. For
example, \cite{FenechConti16} examined the additive bias as a function
of galaxy separation, which is also affected by blending in the
presence of an anisotropic PSF. This can be used to parametrise the
residual biases caused by blending, thus reducing the sensitivity to
the local density.  Hence, in addition to the statistics of the local
background, information about the local galaxy density should be
included in the next generation of shape measurement algorithms.

Naively, an easier solution would be to remove close pairs from the
analysis. Clear cases may indeed be identified and culled from the
data, but some galaxies are blended to such a large degree that they
are detected as single objects. As shown by \cite{Dawson16} the latter
are particularly relevant for deep ground-based observations and lead
to an increase in the shape noise. Importantly, very strict criteria
may result in undesirable reductions in source densities, especially
in the case of deep data. The challenge is thus to find a balance
between the sensitivity of the multiplicative bias due to blending and
the increase in statistical uncertainties when blends are
removed. Moreover, the preferential removal of sources behind
over-dense regions, where the lensing signal is highest, complicates
the interpretation of the cosmological signal, as was shown by
\cite{Hartlap11}.  If ignored, the resulting shear correlation
function can be biased low by a few percent on scales of $1'$. On
the other hand, close pairs may also affect the fidelity of
photometric redshift estimates. The impact of blending on the
combination of shape and redshift determination is another open
question, which requires further study using multi-band image
simulations.

\subsection{Impact on size magnification}
\label{sec:magnification}

So far we have focused on the measurements of galaxy shapes, but
gravitational lensing also alters the sizes of galaxies and their
fluxes because surface brightness is conserved. Although the
measurement of the magnification signal is generally noisier than the
shear, it can in principle be made using the same data that are used
in the shear analysis. Moreover, the shear field is related to the
projected surface density through a convolution \citep[e.g.][]{KS93},
whereas magnification, to leading order, probes the convergence field
directly. This can help break parameter degeneracies, in particular
for studies of density profiles.

The change in flux modifies the number density of sources for a
magnitude limited sample, where the net effect depends on the slope of
the number counts. The sources need not be resolved to measure this
flux magnification signal, thus expanding the sample of potential
sources to be used.  The signal has been measured around clusters of
galaxies \citep[e.g.][]{Hildebrandt11,Ford14,Umetsu16} and galaxies
\citep[e.g.][]{Hildebrandt09}. The main challenges for flux
magnification studies are the need for uniform photometry and a very
clean separation of lens and source samples \citep[see][for a detailed
discussion on observational biases in flux magnification
measurements]{Hildebrandt16a}.

On the other hand, the change in galaxy sizes, or size magnification,
has not been widely used, because it requires high quality imaging. A
number of results have been presented based on HST
observations. \cite{Schmidt12} studied a sample of galaxy groups using
a combination of flux and size magnification, finding fair agreement
between the shear and magnification measurements. \cite{Duncan16} used
HST observations of the A901/A902 supercluster and found that
the magnification measurements yielded lower masses, although the
statistical uncertainties are substantial.

\cite{Casaponsa13} studied how well size magnification can be measured
with {\tt lensfit} \citep{Miller13} and concluded that an unbiased
estimate of the convergence can be obtained, provided the source
galaxies are larger than the PSF and have a SNR$>10$. These
constraints are similar to those for reliable shape
measurements. Their image simulations, however, ignored the impact of
blending, as each postage stamp contained only a single galaxy. As
blending tends to bias the measured sizes, it is worthwhile to explore
this using our more realistic simulations.  The image simulations that
we use to study the performance of shear measurements can be used to
identify additive biases for magnification, as any change in mean size
must be the result of a systematic. However, to quantify the
systematics for size magnification studies in more detail we simulated
the impact of pure magnification, including the changes in flux. To do
so, we magnified the galaxies in the input catalog (including the mean
separation between the galaxies) by a factor
$1+{\cal M}_{\rm mag} \in [1.0, 1.05, 1.1]$ and analysed these images
using our standard pipeline. Our lensing pipeline does not attempt to
estimate PSF-corrected sizes, and instead we use the observed
half-light radii to examine biases in magnification studies.

We take the simulation with ${\cal M}_{\rm mag}=0$ (i.e., no
magnification) and $n_{\rm fac}=1$ as the reference, and compute
\begin{equation}
{\cal M}_{\rm mag}^{\rm obs}=\frac{\langle \tilde{r}_{\rm h}\rangle}
{\langle \tilde{r}_{\rm h}\rangle_{\rm ref}}-1,\label{eq:mag}
\end{equation}
\noindent where $\tilde r_{\rm h}$ is the half-light radius from which
the mean PSF size was subtracted in quadrature. We found that this
simple estimator scales linearly with the input
magnification. Analogous to what was done to quantify the biases in
the shape measurement algorithm, we define
\begin{equation}
{\cal M}_{\rm mag}^{\rm obs}=\mu_{\rm mag}\, {\cal M}^{\rm true}_{\rm mag}+c_{\rm mag},
\end{equation}
\noindent where $\mu_{\rm mag}$ and $c_{\rm mag}$ are the multiplicative
and additive bias, respectively. We note that our estimate of the size
may be particularly sensitive to blending, and a more detailed study
is warranted. Nonetheless with the simulations it is possible to
highlight some of the challenges for magnification studies.

\begin{figure}
\centering
\includegraphics[width=8.5cm]{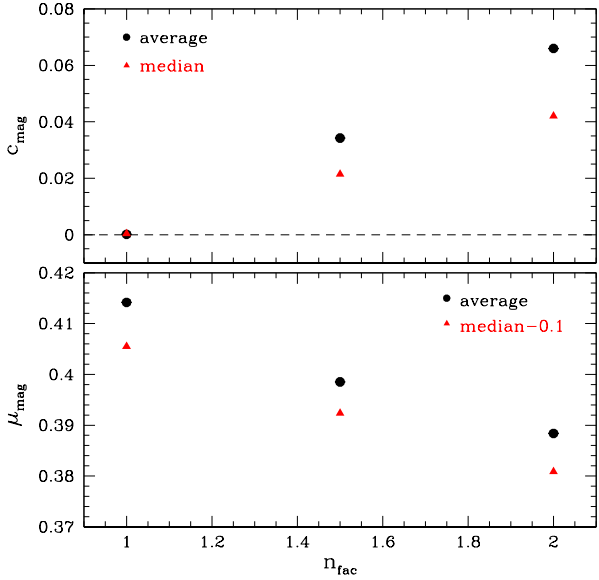}
\caption{Multiplicative (bottom panel) and additive bias (top panel) for 
size magnification, as a function of the relative increase in the number 
density of galaxies brigher than magnitude 26. The measurements are
based on the observed half-light radius of galaxies with magnitude
$20<m<24.5$; the black points use the average size, whereas the
red triangles correspond to the results when the median size is used
(we subtract 0.1 from $\mu_{\rm mag}$ in this case for easier comparison).
Part of the multiplicative bias is the result of smaller 
galaxies entering this magnitude-limited sample due to magnification.
\label{fig:bias_magnify}}
\end{figure}

The points in the bottom panel in Fig.~\ref{fig:bias_magnify} indicate
the multiplicative bias $\mu_{\rm mag}$ for galaxies with $20<m<24.5$
as a function of $n_{\rm fac}$, the increase in source number density
relative to the reference simulation. As was the case for shape
measurements, the multiplicative bias is affected by blending. If we
use the median size instead in Eqn.~(\ref{eq:mag}), we find similar
results but with smaller biases; the red triangles in the bottom panel
of Fig.~\ref{fig:bias_magnify} are offset by 0.1 to allow for a direct
comparison.

The observed change in average size is a combination of the increase
in size due to magnification, and an increase in the number of
intrinsically smaller galaxies due to flux magnification. The latter
is quite relevant: if we repeat the analysis by adjusting the
magnitude limits to correct for the change in flux, we find
$\mu_{\rm mag}=0.66$, and $\mu_{\rm mag}=0.81$ if we use the
median size. The sensitivity to the properties of galaxies below the
nominal flux limit is an additional complication for magnification
studies that is essentially absent in the case of shape measurements.

More worrisome are the results presented in the top panel in
Fig.~\ref{fig:bias_magnify}: the additive bias $c_{\rm mag}$ is a
strong function of the number density of bright galaxies. The bias is
reduced somewhat if we use the median size, as indicated by the red
triangles, and the use of more optimized size estimates may improve
things further. However, as the regions of high magnification tend to
correspond to regions of increased galaxy density, this substantial
additive bias represents a serious complication, especially for
cluster studies such as the one presented in \cite{Duncan16} or
\cite{Schmidt12}.

Our results suggest that magnification studies will also be affected
by the complexity in the data. In particular the additive bias
  that arises from changes in the galaxy density needs to be carefully
  accounted for. Although this is possible in principle, the argument
that magnification is an attractive complement to cosmic shear because
it is subject to different systematics \citep{Alsing15} should be
reconsidered: our findings suggest rather that magnification is
subject to {\it additional} systematics.

\section{The impact of undetected galaxies}
\label{sec:faint}

The properties of sufficiently bright galaxies can be compared
directly to the outputs of the simulations, and remaining trends can
be quantified and accounted for, for instance through machine learning
techniques. H15, however found that galaxies fainter than the limit of
the source sample also affect the multiplicative bias of the brighter
galaxies. This is partly the result of blending, but also because
these galaxies act as a skewed source of background noise, affecting
the local background determination. If such faint galaxies are not
included in the image simulations the multiplicative bias can be
underestimated by a fair amount: H15 found that the bias doubled for
their simulation of ground based data, and saturated when the
simulation included galaxies that were at least 1.5 magnitude fainter
than the limiting magnitude of the source sample.

We therefore examine which value for $m_{\rm lim}$, the limiting
magnitude of the faintest galaxies included in the simulation, may be
adequate for image simulations for \Euclid. The input GEMS catalog is
incomplete for $m>25.5$ (see Fig.~\ref{fig:input_counts}) and we
augment the catalog by duplicating the fainter galaxies such that the
input counts follow the power-law relation seen at brighter
magnitudes, i.e.  we adopt a slope of 0.36 over the full magnitude
range (indicated by the dashed line in
Fig.~\ref{fig:input_counts}). Compared to the actual counts observed
from the UDF by \cite{Coe06}, our results represent a worst-case
situation. We assign sizes following our parametric model described in
Appendix~\ref{app:size}, but explore the sensitivity of the results to
the size distribution in \S\ref{sec:galsize}.

\begin{figure}
\centering
\leavevmode \hbox{%
\includegraphics[width=8.5cm]{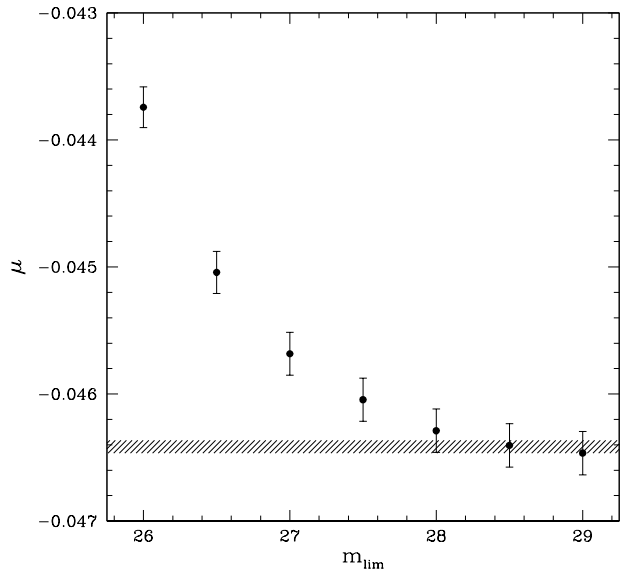}}
\caption{Multiplicative bias for galaxies with $20<m<24.5$ when only 
galaxies with magnitudes brighter than $m_{\rm lim}$ are included in the 
simulation. Because of the small PSF, even galaxies as faint as $m=29$
affect the bias. The hatched region indicates a variation in 
$\mu$ of amplitude $10^{-4}$.
\label{fig:bias_mlim}}
\end{figure}

Figure~\ref{fig:bias_mlim} shows that the multiplicative bias $\mu$
converges rather slowly as a function of $m_{\rm lim}$. To reduce the
number of simulations, we add fainter galaxies to the existing images,
such that the bright galaxies are always in common. Hence the
variation between points is somewhat smaller than the error bars,
which indicate the statistical uncertainty of a single measurement.
The results are based on 81 sets of simulations for each data
point. The dashed line indicates a change $|\Delta\mu|=10^{-4}$,
indicating that we may have to include galaxies as faint as $m=29$ in
the \Euclid~ image simulations.  

The change in bias presented in Fig.~\ref{fig:bias_mlim} is the
result of two effects. Galaxies just below the detection limit
affect the shape measurements mostly through blending, whereas the very
faint galaxies bias the measurements by affecting the determination
of the local background. We revisit this topic in \S\ref{sec:implementation} 
where we explore different implementations of the background determination.

\subsection{Varying size distribution of faint galaxies}
\label{sec:galsize}

The effect of the faint galaxies is to add highly skewed noise to the
pixels where they are located. Our parametrized model for the galaxy
sizes (see Fig.~\ref{fig:model_size}) suggests that the faintest
galaxies are small, but have sizes that are not completely negligible
compared to the size of the PSF. This is different from the large PSF
in ground-based data.

Size distributions for relatively bright galaxies $(m<26$) can be
determined from existing deep HST observations, such as the
Cosmological Evolution Survey \citep[COSMOS;][]{Scoville07} and the
All-wavelength Extended Groth Strip International Survey
\citep[AEGIS;][]{Davis07}. For instance, \cite{Griffith12} present a
compilation of photometric and morphological measurements.  Reliable
size estimates of the fainter galaxies require deeper data, which are
only available for relatively small areas, such as the UDF. 

To determine whether the available data are adequate, we examine how
well the mean size needs to be determined. To do so we increase the
input half-light radii of galaxies with $m>27$ by a factor
$f^{\rm faint}_{\rm size}$ and measure the difference $\Delta\mu$ with
respect to the reference simulation. To reduce the number of
simulations, the positions and intrinsic ellipticities of the galaxies
are the same for the different values of $f^{\rm faint}_{\rm size}$.

\begin{figure}
\centering
\includegraphics[width=8.5cm]{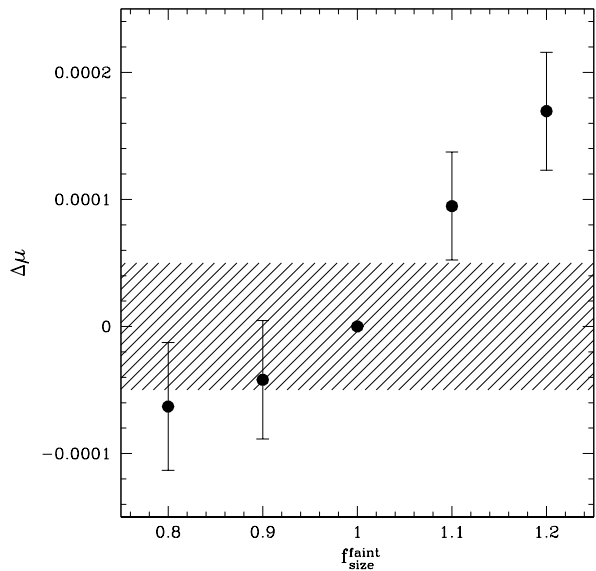}
\caption{Change in multiplicative bias for galaxies
with $20<m<24.5$ when the sizes of the input galaxies
with $m>27$ are increased by a factor $f^{\rm faint}_{\rm size}$ compared
to our reference distribution. The hatched region indicates
$|\Delta\mu|<5\times 10^{-5}$.
\label{fig:bias_size}}
\end{figure}

The results, based on 50 sets of simulations for each value of
$f^{\rm faint}_{\rm size}$, are presented in
Fig.~\ref{fig:bias_size}. We find that the multiplicative bias is
smaller (corresponding to positive $\Delta\mu$ because $\mu<0$) when
the faint galaxies are larger.  This is expected, because the galaxies
are more spread out and thus introduce noise that is less skewed. If
we consider an allocation $|\Delta\mu|<5\times 10^{-5}$, these results
indicate that the mean sizes of galaxies with $m>27$ should be
determined to better than $5\%$.

Given the width of the observed size distribution and the number of
galaxies for which sizes were determined in the UDF, we find that the
mean sizes of these faint galaxies can in principle be constrained to
better than $4\%$. We note, however, that the analysis presented
by \cite{Coe06} will need to be revisited to ensure that biases in the
mean sizes are sufficiently small.

\subsection{Varying the count slope of faint galaxies}
\label{sec:slope}

We expect the amplitude of the multiplicative bias to decrease if
fewer faint galaxies are present. For our reference model we adopted a
single powerlaw slope for the galaxy counts of 0.36 down to magnitude
29. The actual counts from the UDF from \cite{Coe06} shown in
Fig.~\ref{fig:input_counts} suggests that the actual slope is lower;
we obtain a best fit value of $0.237\pm 0.009$ when we fit a powerlaw
to the counts of galaxies with $25<m<29$. The error on the slope was
obtained by splitting the data into four quadrants. To quantify the
sensitivity of our results to the count slope of faint galaxies we
simply change the slope of the counts for galaxies with $m>24.5$ to a
value $\alpha_{\rm faint}$.

\begin{figure}
\centering
\includegraphics[width=8.5cm]{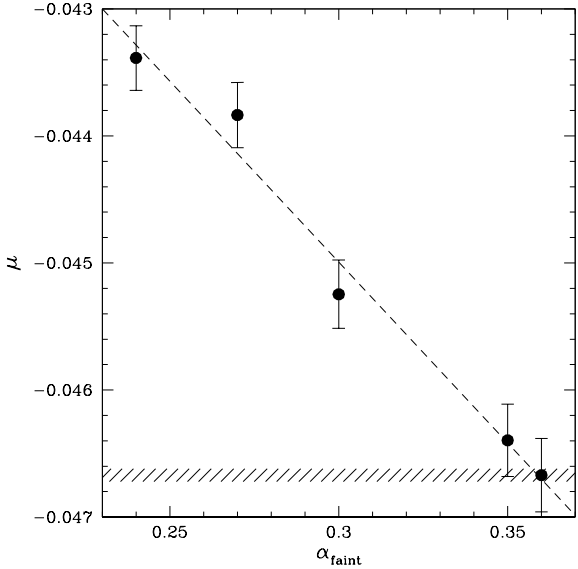}
\caption{Multiplicative bias for galaxies with $20<m<24.5$ as
  function of $\alpha_{\rm faint}$, the powerlaw slope of the galaxies counts 
  fainter than magnitude 24.5; the reference simulation assumes 
  $\alpha_{\rm faint}=0.36$ for all magnitudes, whereas the UDF counts suggest 
  a slope of 0.24 for faint magnitudes. The hatched regions indicates a 
  change of $10^{-4}$ in the estimate of $\mu$.
  \label{fig:bias_slope}}
\end{figure}

The multiplicative bias as a function of $\alpha_{\rm faint}$ is
presented in Fig.~\ref{fig:bias_slope}. The bias increases linearly
with increasing slope. We find a best fit
$d\mu/d\alpha_{\rm faint}=-0.0239\pm0.0014$, which suggests that we need to
determine the mean slope with a precision of $0.004$ if we wish
to allocate a maximum uncertainty of $\Delta\mu=10^{-4}$. This
precision can probably not be achieved from the UDF alone, as we
expect the slope to vary due to variations in the distant large-scale
structure. The impact of cosmic variance can be reduced by combining
with the Hubble Deep Fields, the parallel observations from the
Frontier Fields (although the clusters may contaminate the counts), as
well as future {\it James Webb} Space Telescope observations. These
combined observations should reduce the uncertainties to the required
level.

The multiplicative bias will vary locally as as result of fluctuations
in the faint galaxy counts. If these are uncorrelated with the lensing
signal, the main impact is to slightly increase the noise in the
cosmic shear signal. However, the slope may be affected by
gravitational lensing: as discussed in \S\ref{sec:magnification}
magnification leads to an increase or decrease in the number density
of background galaxies, depending on the slope of the number counts as
a function of magnitude. The relatively flat slope we find for faint
galaxies would lead to a reduction in the average number counts behind
overdense regions. Hence, this introduces a correlation between the
large-scale structure that causes the lensing signal and the
multiplicative bias. This is worrisome, but a detailed assessment is
beyond the scope of this paper. In future work we will explore the
complications that arise from magnification using numerical
simulations to create more realistic source samples.

\section{Varying the input ellipticity distribution}
\label{sec:edist}

The response of a galaxy to an applied shear depends on its
ellipticity: the change in shape is larger for intrinsically round
objects compared to those that are intrinsically elliptical. As the
shear is computed as an ensemble average of the ellipticity of a
population of galaxies, the average shear responsivity depends on the
ellipticity distribution. Moreover, as is discussed in detail in
\cite{Viola14}, the bias due to pixel noise also depends on
the ellipticity distribution.

In this section we examine the sensitivity of the multiplicative bias
to the input ellipticity distribution. H15 found that the
multiplicative bias for their implementation of the KSB algorithm was
relatively insensitive to the value of $\epsilon_0$, because of the
use of a radial weight function (i.e., no attempt is made to optimize
the weight function and match it to the galaxy shape).  Nonetheless
the recovered bias varied by 0.01 over the large range in $\epsilon_0$
considered. Given the much more challenging constraints for \Euclid,
it is interesting to quantify how the bias depends on the ellipticity
distribution used in the image simulations.

\begin{figure}
\centering
\includegraphics[width=8.5cm]{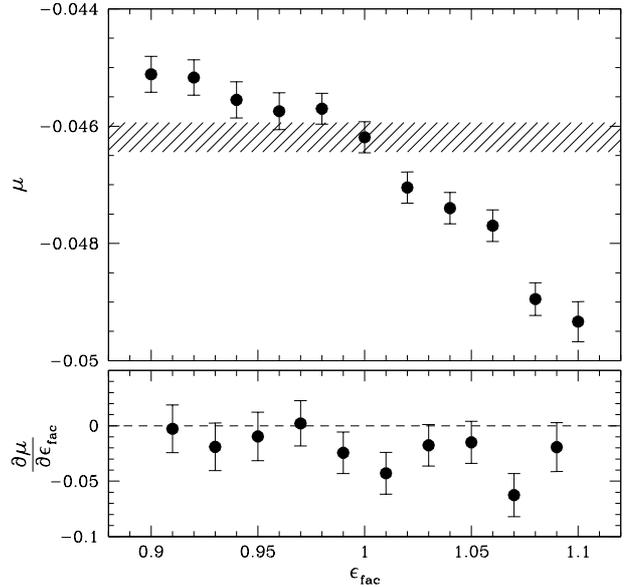}
\caption{{\it Top panel:} Multiplicative bias for  galaxies with $20<m<24.5$
when the input ellipticities are increased by a factor $\epsilon_{\rm fac}$.
The hatched region indicates the allocated uncertainty in the multiplicative 
bias of $5\times 10^{-4}$. {\it Bottom panel:} The slope 
$\partial\mu/\partial\epsilon_{\rm fac}$ as a function of $\epsilon_{\rm fac}$
\label{fig:bias_edist}}
\end{figure}

The input ellipticity distribution can be modified in a number of
ways, but here we simply multiply the input ellipticity by a factor
$\epsilon_{\rm fac}$. The resulting multiplicative bias for galaxies
with $20<m<24.5$ as a function of $\epsilon_{\rm fac}$ is presented in
the top panel of Fig.~\ref{fig:bias_edist}. The hatched region
indicates an uncertainty of $5\times 10^{-4}$ in the multiplicative
bias due to uncertainties in the ellipticity distribution. 

The bottom panel of Fig.~\ref{fig:bias_edist} shows the slope
$\partial\mu/\partial\epsilon_{\rm fac}$ as a function of
$\epsilon_{\rm fac}$, which reaches a maximum absolute value of
$0.05$; for an allocation of $\Delta\mu=5\times 10^{-4}$, this
implies that $\epsilon_{\rm fac}$ needs to be known with a precision
of 1\%. This is comparable to the precision of 0.3 per cent found by
\cite{Viola14} for galaxies with ${\rm SNR}=10$.

Similar to the case for the observed size distribution, a mismatch
between the data and the simulations can be readily identified by
comparing the distributions. A simple comparison of the observed
dispersion $\langle\epsilon_{\rm obs}^2\rangle^{1/2}$ for a subset of
the simulations was sufficient to constrain $\epsilon_{\rm fac}$ to
better than 0.5\%. A complication for real data is that variations in
the noise level will lead to varying broadening of the observed
distributions. Instead \cite{Viola14} advocate observing a small area
of sky to a depth such that the impact of pixel noise can be
neglected. In the case of \Euclid\ an area of $40$ deg$^2$
observed with 40 times the nominal integration time is sufficient to
achieve this. However, further study is needed, as \cite{Viola14} do
not account for the blending with fainter galaxies, which may act as
an additional source of noise.

\section{Sensitivity to implementation}
\label{sec:implementation}

Until now we have focused on the sensitivity of the shape measurement
algorithm to changes in the input images. Although the implementation
of a simple moment-based method such as KSB is relatively
straightforward, a number of choices are made, such as the values of
the {\tt SExtractor} deblending parameters, the width of the weight
function, the background determination, etc. It is therefore useful to
explore how changes in the actual implementation of a method affect
the resulting multiplicative bias. This may help identify parameters
that can improve the robustness. Table~\ref{tab:implement} lists the
multiplicative bias values for simulations with $m_{\rm lim}=29$ for
the various implementation changes that we discuss in more detail
below.

\begin{table}
\begin{center}
  \caption{Multiplicative bias for $m_{\rm lim}=29$ for different
    implementations\label{tab:implement}}
\begin{tabular*}{\columnwidth}{lc}
\hline
\hline
implementation                   & $\mu$ \\
\hline
reference                              & $-0.04645 \pm 0.00017$ \\
unflagged objects                      & $-0.04146\pm 0.00017$  \\
unmasked neighbouring objects          & $-0.04825 \pm 0.00020$ \\
aperture of $3\, r_{\rm g}^{\rm ref}$ & $-0.07972 \pm 0.00016$ \\
$r_{\rm g}=1.5\, r_{\rm rg}^{\rm ref}$  & $-0.03873 \pm 0.00021$ \\
$r_{\rm g}=2\, r_{\rm rg}^{\rm ref}$    & $-0.08172 \pm 0.00023$ \\
$32<r_{\rm bg}<64$ pixels                & $-0.04267 \pm 0.00017$\\
$64<r_{\rm bg}<128$ pixels               & $-0.04246 \pm 0.00017$\\
  zero background                      & $-0.03779 \pm 0.00026$\\ 
{\tt Swarp} background subtracted      & $-0.03858 \pm 0.00031$\\
\hline
\hline
\end{tabular*}
\bigskip
\begin{minipage}{\linewidth}
{\footnotesize Column 1: modified implementation as explained in the text; 
Column 2: value for the multiplicative bias for galaxies with $20<m<24.5$ 
when galaxies down to $m_{\rm lim}=29$ are included in the simulation.}
\end{minipage}
\end{center}
\end{table}

In our reference analysis we consider all galaxies detected by
{\tt SExtractor}, which also provides a flag to indicate potential
problems with an object. We expect that restricting the analysis to
unflagged objects (flag=0) reduces the bias, which is indeed the case. In
contrast, when we do not mask the neighbouring galaxies, the bias is
increased somewhat.

When we reduce the size of the postage stamp to
$3\, r_{\rm g}^{\rm ref}$, the bias increases significantly,
because the small radial extent effectively truncates the estimates of
the higher order moments. The results presented in
Table~\ref{tab:implement} indicate that the biggest improvement is
achieved by increasing the width of the weight function to 1.5 times
the nominal value for $r^{\rm ref}_{\rm g}$. However, increasing the
width further rapidly increases the bias. Although most of the changes
are small, they are nonetheless comparable or larger than the desired
error budget of $\delta\mu=10^{-4}$.

\begin{figure}
\centering
\includegraphics[width=8.5cm]{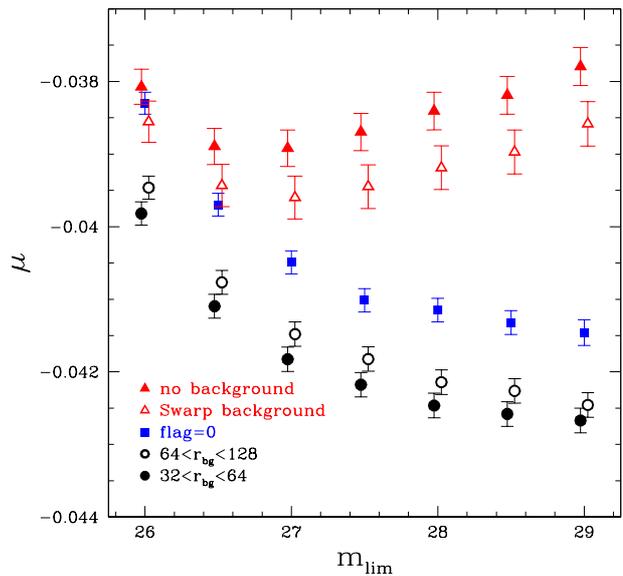}
\caption{Multiplicative bias for galaxies with $20<m<24.5$ as a function of 
$m_{\rm lim}$, the magnitude of the faintest galaxies that are included in the 
simulation, for different implementations as described in the text. In all 
cases faint galaxies affect the measured multiplicative bias. 
\label{fig:bias_mlim_bgvar}}
\end{figure}

As discussed in \S\ref{sec:faint}, faint galaxies affect the local
background determination. The reference analysis determines the
background by fitting a plane to the quadrants of an annulus with
radii $16<r_{\rm bg}<32$ pixels. Increasing the area of this annulus
should make the background estimate more robust. Indeed we find that
the bias is reduced for larger annuli, and find similar values for
both cases.  However, our simulated data have a constant background
with an expectation value that vanishes. Hence we can also consider
the case where the background is set to 0, its true value. This yields
the smallest bias of the changes considered here. To mimic a more
realistic scenario we also run
{\tt Swarp}\footnote{\tt{http://www.astromatic.net/software/swarp}}
with {\tt BACK\_SIZE=128} and {\tt BACK\_FILTERSIZE=3} to determine
the background for each image. This results in a slight increase of
the bias, but it does perform better than our local background
estimates.

Modifying the background determination can thus have a significant
impact on the bias. If the sensitivity to the faint galaxies could be
reduced, then this represents an interesting avenue to improve the
robustness of the shape measurements against variations in the
properties of galaxies that are too faint to be detected in the survey
data. It is therefore interesting to repeat the analysis of the bias
as a function of $m_{\rm lim}$ for different background determinations.  

The results for galaxies with $20<m<24.5$ are presented in
Fig.~\ref{fig:bias_mlim_bgvar}. The solid and open points indicate
the results when we use larger annuli for the local background
determination. The dependence on $m_{\rm lim}$ is, however, similar to
that of the reference setup. We also show results when we reduce the
impact of blending by selecting only galaxies for which no
{\tt SExtractor} flag was raised (blue squares). The bias is indeed
smaller, but it increases for larger values of $m_{\rm lim}$.

The behaviour is markedly different when we assume that we know the
background level a priori (filled red triangles), which does yield the
smallest bias: after initially increasing in amplitude, the
multiplicative bias decreases with increasing $m_{\rm lim}$. This is
also observed when we use {\tt Swarp} to determine the
background in a more realistic way (open red triangles). Compared to
the results for the local background determination, the dependence of
the bias on $m_{\rm lim}$ is in fact stronger.

To examine the origin of these different dependencies we compared the
half-light radii and fluxes measured for galaxies with $20<m<24.5$ in
simulations with $m_{\rm lim}=26$ to measurements obtained using
$m_{\rm lim}=29$, where we matched the two catalogs.  In the case of
local background determination, the inclusion of faint galaxies
results in a small decrease of only $0.2\%$ in the half-light radius,
whereas the flux decreases by $0.3\%$ (both results are median
values). Moreover the distributions of the differences are fairly
symmetric. In contrast fixing the background to zero or using
{\tt Swarp} to subtract a smooth background results in a highly
skewed distribution, where the measurements for most objects are
unchanged. In this case the presence of faint galaxies leads to a
median increase of 2.2\% in size and a median increase of 3.6\% in
flux. This is not surprising, as the faint objects can only increase
the measured sizes and fluxes if the background is fixed. Hence, this
is a generic result, although the details will depend on the shape
measurement method.

The sensitivity to faint galaxies depends clearly on the way the
background is determined. Interestingly a global background
determination may actually lead to an increased sensitivity to
undetected galaxies, although we note that the number density of faint
galaxies in our simulations is higher than observed in deep HST
data (see Fig.~\ref{fig:input_counts}). Although further study is
warranted, we do advocate the use of local background determination,
because it also avoids coherent biases in shapes measurements that may
arise from errors in the global background.

\begin{figure*}
\centering
\includegraphics[width=8.5cm]{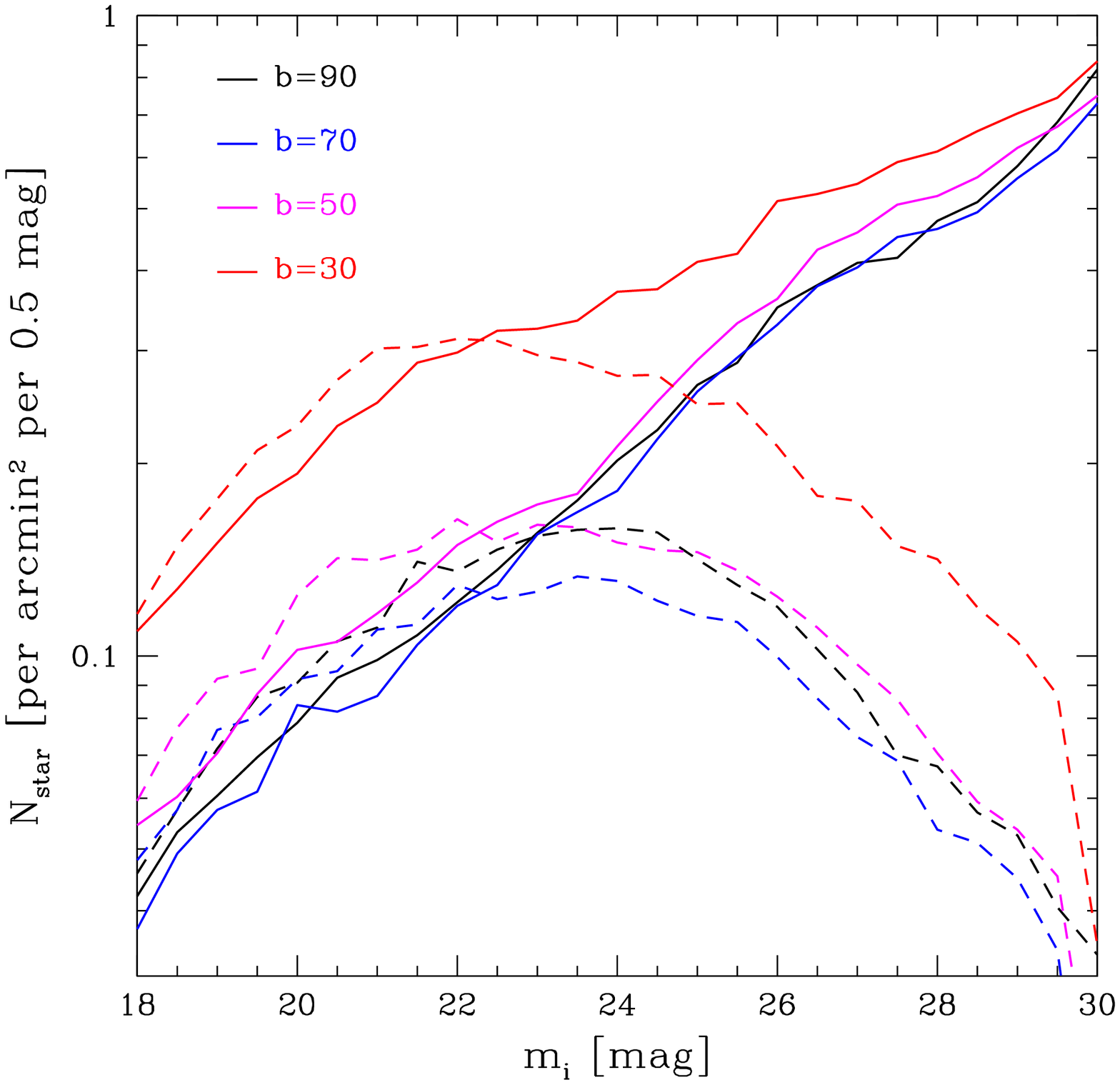}
\includegraphics[width=8.5cm]{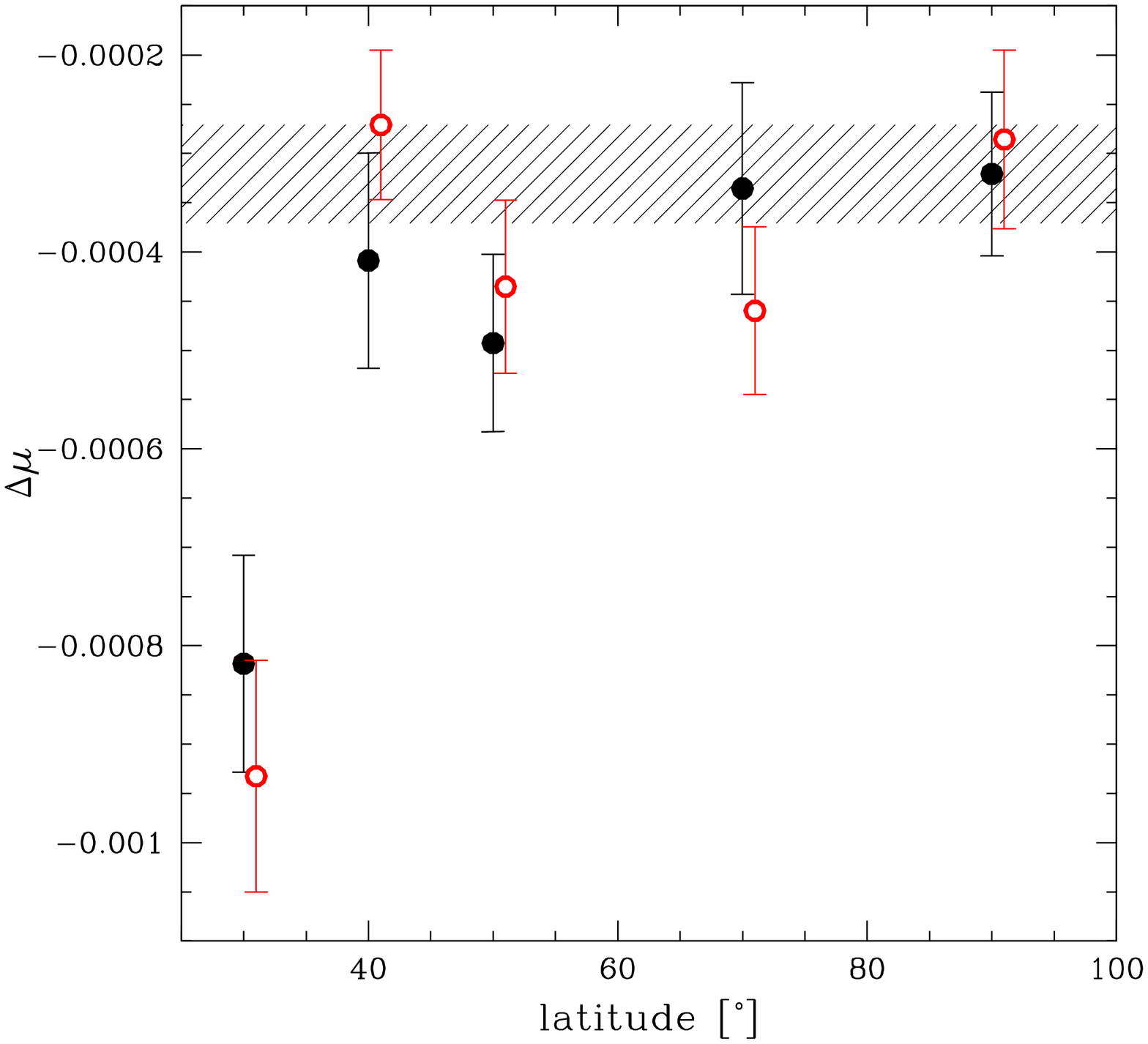}
\caption{{\it Left panel:} Star counts as a function of magnitude for a 
Galactic longitude $l=180^\circ$ and latitudes $b$ of 90, 70, 50 and
30 degrees (black, blue, magenta and red lines, respectively).
The solid lines indicate the results from the Besan\c{c}on model
\citep{Robin03}, whereas the dashed lines correspond to the results
from the TRILEGAL model \citep{Girardi05}. For magnitude $m<23$
the agreement between the models is fair, but we observe
large differences for the faint counts. {\it Right panel:} the
change in multiplicative bias $\Delta\mu$ for galaxies with
$20<m<24.5$, with respect to the reference
simulations, which contain only a small number of stars for PSF modelling.
The black (solid) points correspond to the TRILEGAL counts as a 
function of latitude for $l=180^\circ$, whereas
the red (open) points indicate the results when the Besan\c{c}on model star
counts are used instead. The differences in the faint star counts do
not appear to be important.
\label{fig:starcount}}
\end{figure*}

\section{The impact of stars}
\label{sec:stardens}

Actual data will also contain stars, which may enter the galaxy
catalog or affect the shapes of galaxies through blending. H15 showed
that stars contribute to multiplicative bias for ground based
observations at low Galactic latitude (see their Fig.~A2). Bright
stars can be identified morphologically, because the galaxy images are
resolved, but as faint objects contribute to multiplicative bias we
are most concerned about the impact of faint stars.

To quantiy the impact of blending by stars for \Euclid~ observations
we need predictions for the number density of faint stars as a
function of location on the sky. This requires a population synthesis
model of the Milky Way that can predict the properties of the various
stellar populations that make up the Galaxy. Here we consider two
well-known models for the distribution of stars in the Milky Way, the
Besan\c{c}on model\footnote{http://model.obs-besancon.fr/}
\citep{Robin03} and v1.6 of the TRI-dimensional modeL of thE
GALaxy\footnote{http://stev.oapd.inaf.it/cgi-bin/trilegal}
\citep[TRILEGAL;][]{Girardi05}. Both provide web-based interfaces that
return simulated catalogs. 

The synthesis models use theoretical stellar evolution tracks,
prescriptions for the initial mass function (IMF), star formation
history (SFH), age-metallicity relation, extinction and geometries for
the Galactic components, and a stellar spectral library to predict the
photometry at a given location on the sky. The models consider
separate spatial distributions and star formation histories for the
thin and thick disk, the bulge and the stellar halo, and the
parameters are optimised to match certain observations, such as star
counts, colors or kinematic data. 

We note, however, that the assumption of a smooth stellar halo may be
inadequate, as it is well known that the stellar halo contains
substantial substructure as a result of minor mergers, with the
Sagittarius stream \citep{Ibata94} being the most prominent.  Although
the stream can be clearly identified out to large distances using
multi-color observations \citep[e.g.][]{Pila-Diez15}, it is less
prominent in the counts themselves \citep[see e.g. Fig.~4
from][]{Pila-Diez14}.

The Besan\c{c}on and TRILEGAL models provide a good match to observed
star counts at bright magnitudes. For instance, \cite{Gao13} present a
comparison to SDSS star counts at the north Galactic pole. At fainter
magnitudes the situation is less clear. Deep HST observations provide
some constraints \citep[e.g.][]{Pirzkal05, Stanway08, Pirzkal09}, but
also for these data contamination by galaxies limits the selection at
the faintest magnitudes, although proper motions can be used to improve
the selection \citep[e.g.][]{Kilic05}. 

The Besan\c{c}on model parameters are fixed and we consider counts in
the Megacam $i'$ filter, including stars with maximum distance of
300kpc. The results as a function of longitude $b$ (for a Galactic
latitude of $l=180^\circ$) are presented in Fig.~\ref{fig:starcount}
by the solid lines. The Besan\c{c}on model includes a population of
low-luminosity white dwarfs \citep[see \S{2.5.3} in][]{Robin03} which
dominate the faint counts. Although these may be confused with blue
extragalactic sources, we note that a comparison with
Fig.~\ref{fig:input_counts} shows that faint galaxies outnumber the
stars by two orders of magnitude.

The TRILEGAL interface allows the user to change the model parameters,
such as the IMF or binary fraction, but also the geometry and star
formation histories of the bulge, thin disk, thick disk and stellar
halo components. We use the default settings and the resulting counts
as a function of Megacam $i'$ magnitude are indicated by the dashed
lines in Fig.~\ref{fig:starcount}.  For bright stars ($m<23$) the
agreement with the Besan\c{c}on predictions is fair; the large
differences for the faint counts are caused by the large number
of distant halo white dwarfs in the  Besan\c{c}on model. 
In this regard, we can consider the Besan\c{c}on model as a
worst case scenario for the impact of stars on shape measurements.

The right panel of Fig.~\ref{fig:starcount} show the resulting change
in multiplicative bias (relative to the reference simulations that do
not include many stars) as a function of Galactic latitude (for a
Galactic longitude $l=180^\circ$). As for the reference case, we
select galaxies requiring that the observed half-light radius
$r_{\rm h}>0\farcs11$, because the galaxies with $m<24.5$ are larger than
the stars. The black (solid) points indicate the results for the
TRILEGAL star counts, whereas the red (open) points correspond to the
results when using the Besan\c{c}on predictions. We observe a small
increase in multiplicative bias of $3\times 10^{-4}$, which
increases to $10^{-3}$ for low Galactic latitudes. Importantly,
the biases for the two models are consistent, suggesting that the
differences in faint star counts have a negligible impact.

\begin{figure}
\centering
\includegraphics[width=8.5cm]{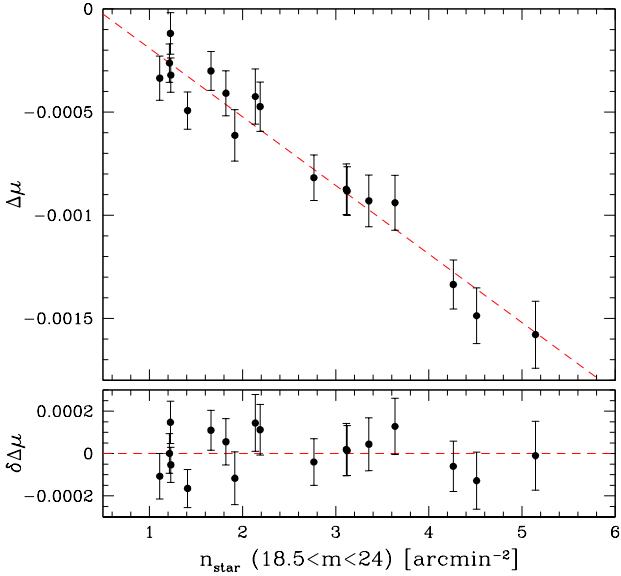}
\caption{{\it Top panel:} Change in multiplicative bias, relative
to the reference simulations (which contain only a small number of stars), 
for galaxies with $20<m<24.5$ as a function of the observed density of 
bright stars ($18.5<m<24$). The red dashed line is the 
best fit linear relation. {\it Bottom panel:} the residuals 
$\delta\Delta\mu$ with respect to this fit, indicating that the bias 
can be predicted with a precision of $10^{-4}$ (rms).
\label{fig:bias_nstar}}
\end{figure}

As the number density of stars varies across the sky, so will the
corresponding change in multiplicative bias. One would like to avoid
having to create large suites of image simulations to quantify the
impact of stars in the data. We therefore explored whether it is
possible to predict the change in multiplicative bias using the
observed counts of bright stars. To do so we used TRILEGAL star counts
for a range of Galactic longitudes and latitudes and determined the
resulting multiplicative bias as a function of $n_{\rm star}$, the
observed number density of stars within the magnitude range
$18.5<m<24$. 

Figure~\ref{fig:bias_nstar} shows that the bias increases linearly
with $n_{\rm star}$. The residuals with respect to the best fit linear
relation are presented in the bottom panel; with an rms of
$10^{-4}$ these results suggest that the bias can be predicted
sufficiently well using the observed number density of stars. Note
that we did not attempt to optimise the magnitude range for the bright
star counts, but further refinement is limited in any case by the
relative small number of simulations we created (20 sets per
position).

\section{Conclusions}

To exploit the statistical power of future cosmic shear surveys, the
accuracy with which shapes can be measured needs to be improved. This
requires a careful modeling of the PSF, which is the dominant source
of bias, even for space-based observations. However, the presence of
noise and neighbouring galaxies affects the accuracy of shape
measurement algorithms.

The performance of shape measurement algorithms can, however, be
quantified using simulated data. Moreover, the resulting biases as a
function of observed properties can be used to derive an empirical
calibration of the method \citep[e.g.][]{Hoekstra15,FenechConti16}.  For
a meaningful calibration, it is essential that the simulated data are
sufficiently realistic, whereas many studies have focused on
idealised cases, which not only yield incorrect bias estimates, but
can also lead to the development of algorithms that may not perform
well on real data. 

We explored the sensitivity of the multiplicative bias of a simple
moment-based method to a number of real-life effects, such as noise,
the presence of stars, and galaxies below the detection limit. We also
examined how well the sizes and intrinsic ellipticities of the
galaxies need to be known. Moreover, we demonstrated how slight
modifications in the algorithm can significantly change the recovered
biases. Most of the changes are well below the statistical
uncertainties of ongoing surveys, but need to be studied for the next
generation of surveys that are an order of magnitude larger.  Hence,
although we focused here on the accuracy with which \Euclid\ will need
to determine galaxy shapes, our findings are also of relevance for
LSST and WFIRST.

We used the publicly  available code {\tt GalSim} \citep{Rowe15} to create
the simulated data. The images are analysed with the KSB algorithm
\citep{KSB95}, which is fast but not well-suited for the analysis of
real data. However, the objective of this paper is not to calibrate
this algorithm, but rather to demonstrate that shape measurements are
sensitive to a number of real-life effects that have not been
considered before. Although the actual sensitivities will differ
between algorithms and implementations, our results are generic.

As expected, the multiplicative bias depends on the noise level in the
images. The noise level will vary across the survey, in particular due
to changes in the zodiacal light, but we find that this can be
measured with sufficient precision from the \Euclid\ data. The bias is
also a strong function of galaxy size, and changes in the input sizes
thus modify the results. This can be accounted for by adjusting the
input parameters by comparing the simulated data to the observations
\citep[e.g.][]{Bruderer16, FenechConti16}, or by capturing the
dependence on size through an empirical correction. It is, however,
important that selection biases are also accounted for
\citep{FenechConti16}.

The multiplicative bias also depends on the galaxy number density,
which is not captured in simulations of individual galaxies. As the
cosmic shear signal is correlated with the galaxy density, it is
important that shape measurement algorithms account for this. For
instance, one could consider algorithms that are minimally sensitive
to neighbouring objects. We find that adjustments in the
implementation can indeed reduce the biases somewhat.  Compared to
fitting algorithms, moment-based methods are more sensitive to blends
because they bias the estimates of the moments.  Whether this can be
reduced will require further study. We also note that, strict criteria
may result in undesirable reductions in source density, but can also
complicate the interpretation of the signal. The number density of
stars also varies across the survey, but we find that the observed
star counts can be used to empirically correct the bias with adequate
accuracy. Blending also affects the accuracy of measurements of the
change in galaxy sizes due to magnification. This results in additive
biases for size magnification studies that will need to be accounted
for.

The importance of galaxies below the detection limit was already
highlighted in H15. Given the accuracy required for \Euclid\ we find
that we need to include galaxies as faint as 29th magnitude. The bias
is sensitive to the number density of these faint galaxies and their
size distribution.  Magnification of faint galaxies by the intervening
large-scale structure may introduce a correlation between the lensing
signal and the multiplicative bias. This requires further study, but
we note that our results may be considered a worst case, because the
observed number density of faint galaxies is much lower than we
simulate (see Fig.~\ref{fig:input_counts}). Minimizing the sensivity
to these faint galaxies is nonetheless an important topic for further
study.

The shape measurement also depends on the way the background level is
determined. Our reference implementation uses a local measurement,
which has the advantage that errors in the background due to artifacts
are confined to small scales. In this case the multiplicative bias
increases as fainter galaxies are included in the simulations. We
observe the opposite, but steeper, trend when we consider a global
background determination. 

We considered a number of complications that occur in the analysis of
real data, and demonstrated that these affect the accuracy of shape
measurements at a level that is relevant for future surveys such as
\Euclid. These should be considered in the selection of shape
measurement algorithms. Improving the accuracy of the input parameters
of the image simulations is perhaps the most important step.  Although
our results show that achieving sub-percent accuracy is challenging,
we estimate that existing HST observations, as well as future
{\it JWST} data, are in principle adequate to create sufficiently 
realistic populations of galaxies.

\vspace{0.5cm} 

We thank the members of the \Euclid\ weak lensing science working
group, in particular Mark Cropper, Tom Kitching, Lance Miller and Tim
Schrabback, for useful discussions. HH also acknowledges fruitful
discussions during an international team meeting on ``Cosmology with
size and flux magnification'' led by Alan Heavens and Hendrik
Hildebrandt at the International Space Science Institute (ISSI). HH,
RH and MV acknowledge support from the European Research Council FP7
grant number 279396. MV is supported by the Netherlands Organisation
for Scientific Research (NWO) through grant 614.001.103.

\bibliographystyle{mnras}
\bibliography{sim}

\appendix

\section{Parametric model of galaxy sizes}
\label{app:size}

To describe the size distribution of faint galaxies we extrapolate the
observed distribution of half-light radii from GEMS \citep{Rix04} to
fainter magnitudes. In this Appendix we describe how we determined the
model parameters.

\begin{figure}
\centering
\leavevmode \hbox{%
\includegraphics[width=8.5cm]{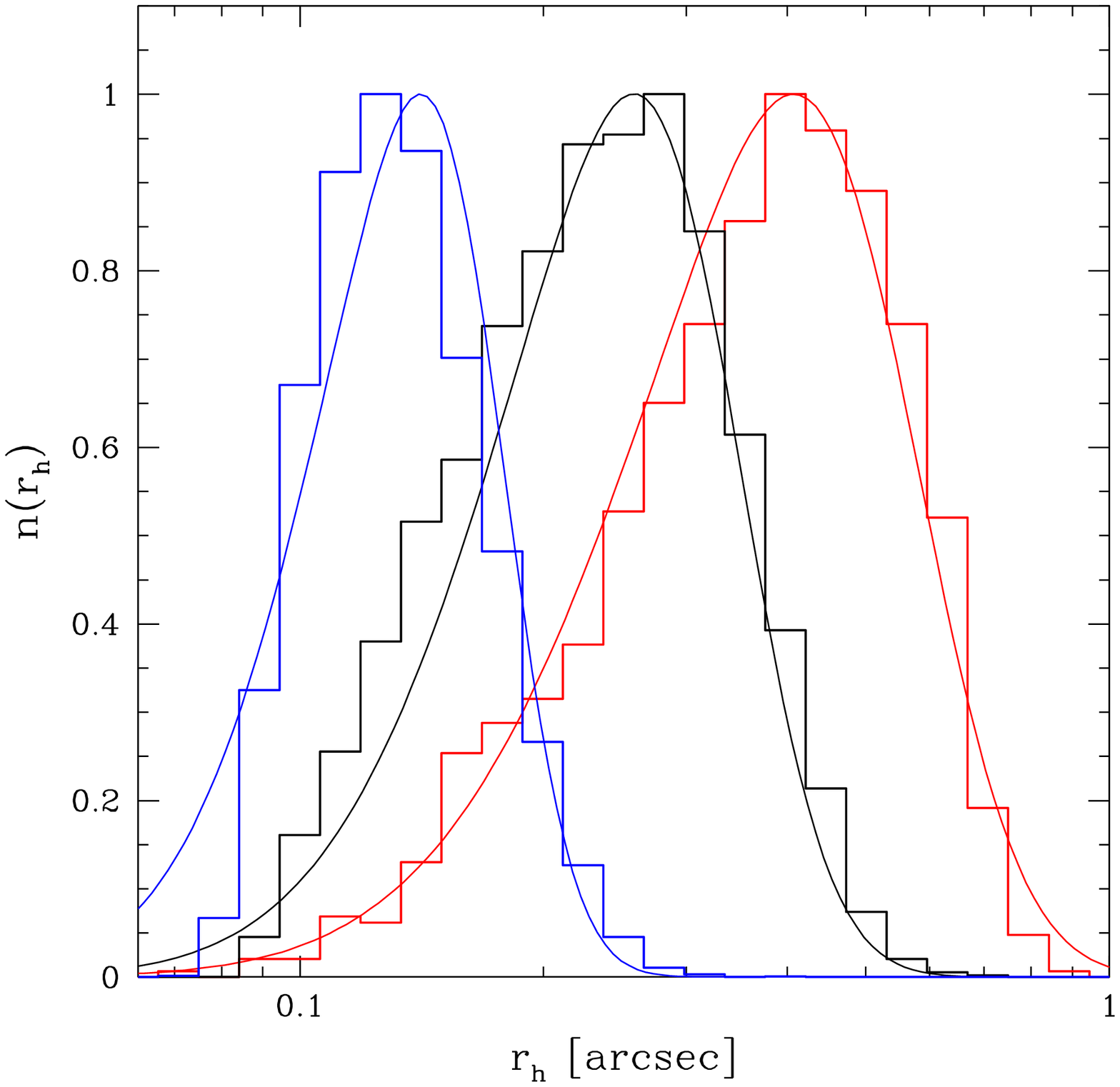}}
\caption{Histograms of the distribution of half-light radii from the GEMS
catalog \citep{Rix04} for three magnitude bins (with width 0.5 magnitude).
From left to right the central values are 26.5, 24.5 and 23.0 (blue, black 
and red, respectively). The smooth curves show the parametric model, which 
match the data fairly well.
\label{fig:size_dist}}
\end{figure}

The observed distribution of half-light radii is well described by a
log-normal distribution, as is shown in Fig.~\ref{fig:size_dist}.
The histograms show the data for three bins with a width of 0.5
magnitude. From left to right the central values are 26.5, 24.5 and
23.0 (blue, black and red, respectively), clearly showing that the
mean galaxy sizes decrease with magnitude. We note that the faintest
bin in Fig.~\ref{fig:size_dist} is shown for reference because the
GEMS catalog is incomplete for this bin. 

Instead we derive the model parameters using bins with central values
between magnitude 23 and 25.5 and assume that we can extrapolate the
results to fainter magnitudes. We initially kept the skewness as a free
parameter, but found that the most robust results were obtained by
fixing it to a representative value of $-0.58$.  In this
magnitude range the mean size and dispersion decrease approximately
linearly with apparent magnitude $m$ (see
Fig.~\ref{fig:model_size}), and we obtain best fits
$\langle\log_{\rm 10} r_{\rm h}\rangle=3.086-0.145\times m$ (with $r_{\rm h}$
in units of arcseconds) and rms
$\sigma_{\log_{10} r_{\rm h}}=0.892-0.0266\times m$. These relations were
used to compute the dashed lines in Fig.~\ref{fig:model_size}, which
indeed match the GEMS data well. The smooth curves in
Fig.~\ref{fig:size_dist} are the model size distributions, which
match the histograms well, including the magnitude 26.5 bin, which is an
extrapolation using the model. We use this simple parametric model to
assign sizes to galaxies $m>26.5$ in our image simulations.

\begin{figure}
\centering
\leavevmode \hbox{%
\includegraphics[width=8.5cm]{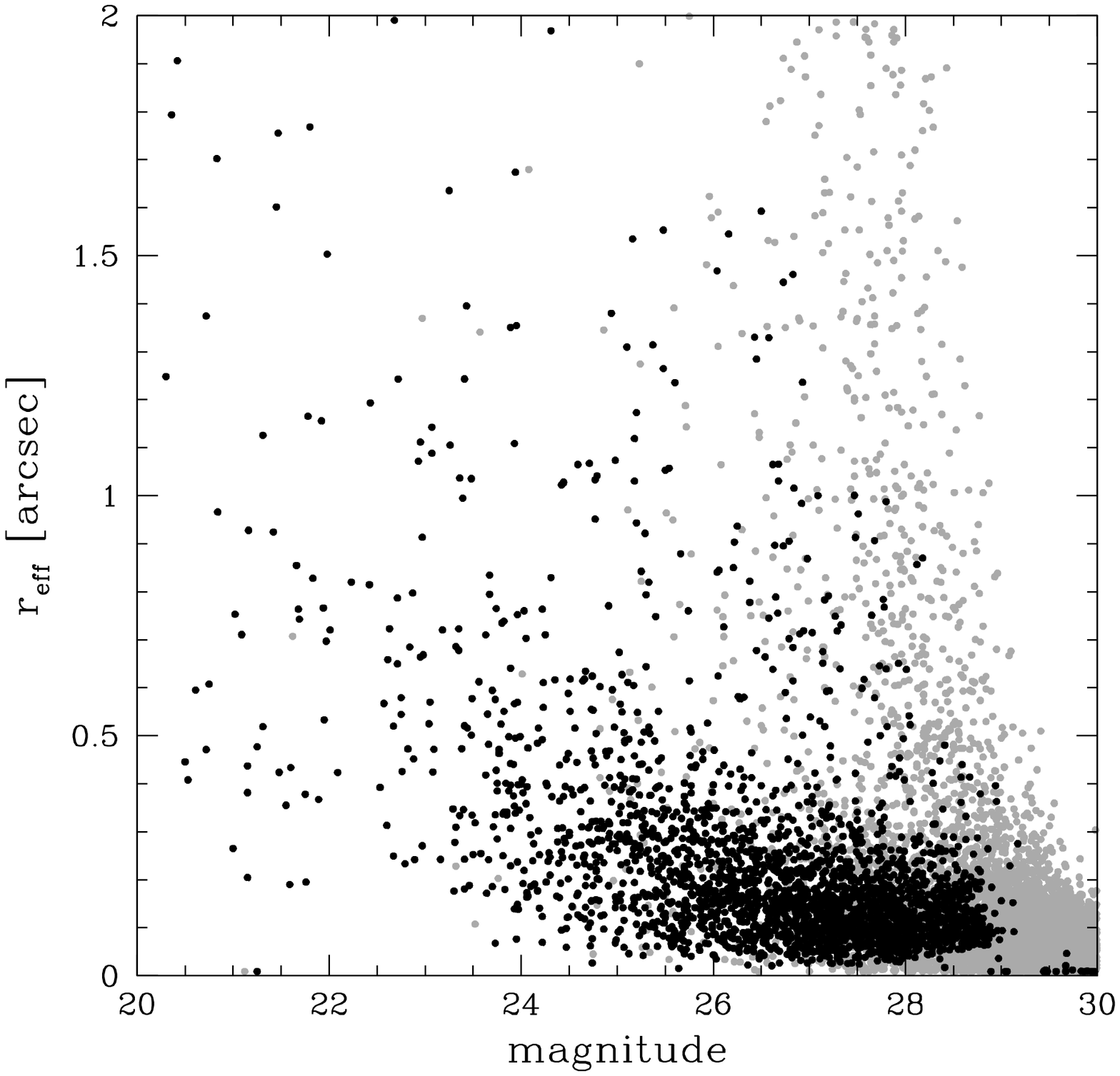}}
\caption{Plot of the best fit effective radius as a function of apparent 
magnitude from \citet{Coe06}. The black points indicate the points with a 
fractional uncertainty $<10\%$ in the size, whereas the grey points 
correspond to the remaining galaxies in the catalog. 
\label{fig:coe}}
\end{figure}

To compare our model to deep HST observations, we use measurements
from the UDF by \cite{Coe06}. Similar to \cite{Rix04} they fit the
galaxy brightness distribution using
{\tt GALFIT}. Figure~\ref{fig:coe} shows the resulting best fit
effective radii as a function of apparent magnitude. The scatter
increases rapidly for galaxies with $27.5<m<29$, most likely due to
degeneracies with other fit parameters. If we require a relative
uncertainty $<10\%$ in $r_{\rm eff}$, the scatter is reduced
considerably, as indicated by the black points in
Fig.~\ref{fig:coe}.

Unfortuntately \cite{Coe06} only report the best fit effective radii,
whereas we use the more robust half-light radius. Instead we assume a
linear relation between $r_{\rm eff}$ and $r_{\rm h}$, and determine
the parameters from the GEMS catalog which contains both size
estimates \citep{Rix04}:
$r_{\rm h}\approx 0.066+0.46\times r_{\rm eff}$. The resulting average
half-light radii are indicated by the red points in
Fig.~\ref{fig:model_size}, which indicate that our parametric model
may underestimate the sizes for the faintest galaxies. Similarly we
adjust the dispersions; the red points in the bottom panel of
Fig.~\ref{fig:model_size} agree fairly well, although the width of the
size distribution may be larger than adopted.

\end{document}

%% file: journals.tex
\def\aj{{AJ}}			
\def\apj{{ApJ}}			
\def\apjl{{ApJ}}		
\def\apjs{{ApJS}}		
\def\aap{{A\&A}}		
\def\aaps{{A\&AS}}		
\def\mnras{{MNRAS}}		
\def\prd{{Phys.~Rev.~D}}	
\def\ssr{{Space~Sci.~Rev.}}	
\def\nat{{Nature}}		

\def\physrep{{Phys.~Rep.}}   